%% file: main.tex
\documentclass[11pt,a4paper]{article}
\pdfoutput=1
\usepackage{ifpdf}
\usepackage{jheppub}
\usepackage{rotating}
\usepackage[bb=boondox]{mathalfa}
\usepackage{comment}
\usepackage{slashed}
\usepackage{tikz-cd}
\usepackage{mathrsfs}
\usepackage{amsmath,amsfonts,amssymb}
\usepackage{mathtools}
\usepackage{dcolumn}
\usepackage{bm}
\usepackage[mathlines]{lineno}
\usepackage{float}
\usepackage{blindtext}
\usepackage{titlesec}
\title{Sections and Chapters}
\usepackage[toc,page]{appendix}
\usepackage{graphicx}
\graphicspath{ {./images/} }
\usepackage[thinc]{esdiff}
\usepackage[euler]{textgreek}
\usepackage{braket}
\usepackage{physics}
\usepackage{xfrac}
\usepackage{soul}
\usepackage{stmaryrd}
\usepackage{pifont}
\usepackage{slashed}
\usepackage{comment}
\usepackage{orcidlink}
\usepackage{multirow}
\usepackage{booktabs}
\usepackage{comment}

\newcommand{\nn}{\nonumber}

\usepackage[normalem]{ulem}
\definecolor{darkgreen}{cmyk}{1,0,1,0.4}
\definecolor{darkcyan}{cmyk}{1,0,0,0.4}

\def\dis{\displaystyle}
\def\barr{\begin{array}}
\def\earr{\end{array}}

\allowdisplaybreaks

\title{\textcolor{black}{From Thermal History to Multi-Messenger Signatures in $\mathbb{Z}_3$ Symmetric Dark Sector}}

\author[a]{Debajyoti Choudhury,\,\orcidlink{0000-0002-8124-0043}\,}
\author[b]{Jaydeb Das,\,\orcidlink{0000-0001-6335-9377}\,}
\author[c]{Divya Sachdeva\,\orcidlink{0000-0003-4318-6733}\,}
\affiliation[a]{Department of Physics and Astrophysics, University of
  Delhi, Delhi-110007, India.}  \affiliation[b]{Department of Physics,
  Indian Institute of Technology Guwahati, Assam-781039, India.}
\affiliation[c]{Department of Physics, Indian Institute of Technology
  Hyderabad, Telangana-502285, India.}
\emailAdd{debchou.physics@gmail.com}
\emailAdd{jaydebphys@rnd.iitg.ac.in}
\emailAdd{divyasachdeva@phy.iith.ac.in} \abstract{We investigate a
  $\mathbb{Z}_3$-symmetric extension of the Standard Model consisting
  of a right-handed neutrino ($N_R$), a dark fermion ($\chi$)
  that dominates the relic density
  and a dark complex scalar ($S$) that facilitates
 a strong first-order electroweak phase
  transition (SFOEWPT). The observed relic abundance,
  is
  achieved through the combined effects of annihilation,
  semi-annihilation, and dark-sector conversion processes over a broad
  region of parameter space consistent with a SFOEWPT. We further
  investigate the resulting multi-messenger signatures, including
  loop-induced direct detection, indirect detection through gamma-ray
  observations, and gravitational wave signals, with the latter lying
  within the projected sensitivities of future space-based detectors
  such as LISA, BBO, and DECIGO. }

\begin{document}
\maketitle
\flushbottom

\preprint{}

\usetikzlibrary{decorations.markings,decorations.pathmorphing}

\tikzset{
fermion/.style={
postaction={decorate},
decoration={markings,mark=at position 0.55 with {\arrow{>}}}
},
antifermion/.style={
postaction={decorate},
decoration={markings,mark=at position 0.55 with {\arrow{<}}}
}
}

\section{Introduction}
Particulate dark matter (DM) is a well-motivated hypothesis to address
various astrophysical and cosmological observations, and over the last
few decades, has been studied extensively. The
simplest scenarios would call for the DM particle to be a color- and
charge-neutral massive particle. With the Standard Model (SM)
neutrinos being disallowed as candidates on account of their being an
impediment to large scale structure formation, a minimal approach is
to add a new particle and impose a symmetry that stabilizes it. A
portal coupling ($g$) is typically introduced to connect the DM
particle with the SM sector. The ensuing interactions determine not
only the DM number density in the early Universe (and inherited by the
present epoch), but also the prospects of its being detected in a
multitude of experiments, whether these be direct detection (wherein
the DM would interact with the ordinary matter in a dedicated
detector), indirect detection (wherein the DM particles annihilate
into SM ones leaving a tell-tale signature) or even collider
production of DM (essentially the opposite of the indirect detection
process).  It stands to reason that for the correct relic abundance to
be obtained through thermal freeze-out, $g$ cannot be arbitrarily
small. On the other hand, non observation at direct detection
experiments such as LUX-ZAPELIN (LZ)~\cite{LZ:2024zvo} or
XENONnT~\cite{XENON:2024wpa} continue to push the upper limits on the
dark matter-nucleon scattering cross section to increasingly smaller
values. Indeed, the lack of an observable signal has not only ruled
out large regions of the parameter space of minimal portal models, but
is increasingly calling into question the very viability of such a
paradigm.

A possible way to evade the bounds from direct detection would be to
consider a $\mathbb{Z}_3$ symmetry (instead of the usual
$\mathbb{Z}_2$) as the one responsible for the DM's
stability. Notionally, this would require three DM (or DM-like)
particles at a vertex, thereby suppressing the rates for direct
detection. The simplest viable models, however, are almost bereft of
any discernible signatures and, thereby, are
uninteresting. Furthermore, minimal Higgs-portal $\mathbb{Z}_2$ and
$\mathbb{Z}_3$ scalar DM models are increasingly constrained when the
observed relic abundance, direct detection limits, vacuum stability
and perturbativity up to the Planck scale are simultaneously
imposed~\cite{Athron_2018}.

An alternative paradigm is to consider an extended dark sector with
additional particles and interactions. Such a framework modifies the
DM phenomenology while relaxing the stringent constraints, and also
opens the possibility that the new particle content simultaneously
addresses other shortcomings of the SM. While
  this may seem an ad hoc complication, it actually is quite
analogous to the SM, where no single parameter is responsible for all
observed phenomena.

Once one moves beyond minimal DM constructions and introduces
additional particles, it is natural to ask whether the new states can
also address other shortcomings of the SM. For example, the
observation of neutrino oscillations has firmly established that
neutrinos possess non-zero masses, nominally requiring physics beyond
the SM. Similarly, baryogenesis requires, amongst other things, an
out-of-equilibrium phase in the thermal history of the Universe, and
is facilitated if the same is actually a strong first-order phase
transition (SFOPT). Within the SM, the electroweak
phase transition (EWPT) is, however, only a smooth
crossover~\cite{Kajantie:1996mn,Csikor:1998eu,DOnofrio:2015gop} and
therefore cannot provide the out-of-equilibrium conditions required
for electroweak baryogenesis.

Traditionally, these problems are often addressed independently. New
fermions are introduced to explain neutrino masses, additional scalar
fields are invoked to modify the EWPT, and a
separate DM sector is added to account for the observed relic
abundance. While such constructions are perfectly viable, they often
introduce several disconnected sectors whose only common feature is
that they lie beyond the SM. 

In this work, we explore the alternative possibility that these
phenomena are not independent. The particles and interactions
responsible for one observation may simultaneously influence the
others. Motivated by DM, neutrino masses, and the possibility of a
strong first-order electroweak phase transition (SFOEWPT), we adopt a bottom-up approach and
introduce a dark-sector framework in which the same particle content
plays multiple roles in the thermal history of the Universe. Consequently, the dark-sector structure affects the
relic abundance, gravitational wave (GW) signals, neutrino phenomenology,
as well as astrophysical observations. Measurements from direct and indirect DM searches, collider experiments, neutrino physics, and GW
observations therefore become complementary probes of the same
framework. We then investigate whether the proposed particle content
and interactions remain consistent with the available multi-messenger
observations and experimental constraints.

The framework considered here is primarily phenomenological. Several
studies addressing a combination of these issues have been discussed
in the
literature~\cite{Ghorbani:2018yfr,Kang:2017mkl,Baker_2017,Choudhury:2026laq,Das:2026zuo,Srivastava:2025oer,Chaudhuri:2022sis,Lu:2026byr,Borah:2024emz,Chiang:2019oms,Borah:2023zsb}. However,
whether it originates from a deeper underlying principle remains an
open question that we leave for future investigation.

This paper is organized as follows. In Sec.\,\ref{section2:Model}, we
introduce the model, its particle content, and the underlying
symmetries. We discuss the scalar and fermionic sectors, the origin of
neutrino masses, and the interactions relevant for DM and
phase transition phenomenology. In Sec.\,\ref{section3:pt}, we study
the thermal history of the model and the role played by the different
particles in the early Universe. Particular emphasis is placed on
finite-temperature effects and the resulting evolution of the dark
sector. After identifying parameter regions consistent with the
observed DM relic abundance and a SFOEWPT, we turn in Sec.\,\ref{section4:signatures} to the
corresponding multi-messenger signatures. We finally summarize our
results and conclude in Sec.\,\ref{section5:conclusion}.

\input{model}

\input{thermalHistory4}

 \input{signatures}

 \input{conclusion}
\input{appendix}

\label{Bibliography}
\bibliographystyle{JHEP}
\bibliography{Refs}

\end{document}

%% file: model.tex
\section{The model}
\label{section2:Model}
We begin by listing the only new fields in our
model.  A natural way to explain the tiny neutrino masses is the
seesaw mechanism, and, to this end, we introduce right-handed
neutrinos (RHNs) $N_{\alpha R}$. While it is natural to have three of
these, phenomenologically only the mass-square differences are known
and one of the light neutrinos could yet be massless. In other words,
it can suffice to have only two of $N_{\alpha R}$. We remain agnostic
as to whether there are two or three.  Together with the SM fields,
these comprise the extended ``standard sector''. In addition, we
introduce a Dirac fermion $\chi$ and a complex scalar $S$, which constitute the dark sector through their non-trivial
transformation under the imposed $\mathbb{Z}_3$ symmetry. While,
naively, could
  also contribute to the DM relic density, its role would be seen to be more
  important in both strengthening the phase transition as well as in providing a portal for DM interactions with
the standard sector. The stability of the DM is ensured by
the aforementioned $\mathbb{Z}_3$ symmetry
under which all the standard sector fields (including the $N_{\alpha
  R}$) are invariant, while the only nontrivial transformations are
defined by
\begin{equation}
\chi \rightarrow e^{i2\pi/3}\chi,
\qquad
S \rightarrow e^{i2\pi/3}S.
\end{equation}
\
The complete $\mathbb{Z}_3$-invariant Lagrangian
can, then, be conveniently parametrized as
\begin{equation}
\mathcal{L} = \mathcal{L}_{\rm SM} +\mathcal{L}_N + \mathcal{L}_{\rm dark}
\end{equation}
where
\begin{equation}
  \dis
  \mathcal{L}_N = \sum_\alpha\overline{N_{\alpha R}} i\slashed{\partial} N_{\alpha R}
- \sum_{\alpha, \beta}\left(
\frac{1}{2} m_{R\alpha\beta} \overline{N_{\alpha R}^c}N_{\beta R} + {\rm h.c.}
\right)
+ \sum_{\alpha=e,\mu,\tau}\sum_{i=1}^3(y_N)_{i \alpha} \bar{L}_i \widetilde{H}N_{R\alpha} + {\rm h.c.}.
\end{equation}
After electroweak symmetry breaking, the neutrino mass
matrix in the $(\nu_L^c\, N_R)^T$ basis takes the form
\begin{equation}
\mathcal{M}_\nu=
\begin{pmatrix}
0 & m_D \\
m_D^T & m_R
\end{pmatrix},
\qquad
(m_D)_{i \alpha} =\frac{v_h}{\sqrt{2}} \, (y_N)_{i \alpha} . 
\end{equation}
The standard type-I see-saw mechanism, then, leads to an
effective neutrino mass matrix 
\begin{equation}
m_\nu \simeq - m_D m_R^{-1} m_D^T.
\end{equation}
Here, $m_D$ denotes a $3 \times 2$ (or $3 \times 3$, as the case may
be) Dirac mass matrix, resulting in a $5 \times 5$ (or $6 \times 6$)
complex symmetric neutrino mass matrix $\mathcal{M}_\nu$.  For RHN
masses around the tens of GeV scale, the observed neutrino masses and oscillation data,
\begin{equation}
\Delta m_{21}^2 \simeq 7.4\times10^{-5}~{\rm eV}^2,
\qquad
|\Delta m_{31}^2| \simeq 2.5\times10^{-3}~{\rm eV}^2,
\end{equation}
can be reproduced with Yukawa couplings $y_N \sim 10^{-8}-10^{-7}$. As
mentioned above, at least two RHN are required to fully reproduce the
observed neutrino oscillation data. However, since this is not our
main goal, henceforth we consider only a single RHN. Indeed, for the
twin purposes of studying the dark sector dynamics and phase
transition phenomenology, it suffices to concentrate on a
representative RHN state that dominantly couples to the dark
sector. The remaining RHNs would, then, have relatively minor impact
on the thermal evolution and dark matter phenomenology. All that we
need from the light neutrino phenomenology is the typical order the
Yukawa couplings $(y_N)_{i \alpha}$.

 The dark sector is governed by the Lagrangian
\begin{equation}
  \mathcal{L}_{\rm dark}
  =  \bar{\chi}(i\slashed{\partial}-m_\chi)\chi + (\partial \mu S)^* (\partial^\mu S) -\Big(
y_1 S\,\overline{\chi^c}\chi
+y_{2} S\,\overline{\chi}N_{R}
+y_{3} S^\dagger \overline{\chi^c}N_{R}
+{\rm h.c.}
\Big)
- V(H,S),
\end{equation}
where, as hinted above, we have limited ourselves to a single
$N_R$. The Yukawa couplings $y_{1,2,3}$ would turn out to have very
crucial roles. The scalar potential for the
SM Higgs doublet $H$ and the dark scalar $S$ is given by
\begin{eqnarray}
V(H,S)
&=&
-\mu_h^2\, H^\dagger H
+\lambda_h\, (H^\dagger H)^2
+\mu_s^2 \,S^\dagger S
+\lambda_s\, (S^\dagger S)^2
\nn\\
&&
+\lambda_{hs}\,(S^\dagger S)(H^\dagger H)
+\frac{\sqrt{2}\mu_3}{3}\left(S^3+{\rm h.c.}\right) \ ,
\label{tree}
\end{eqnarray}
where the first two terms should rightfully be a part of
$\mathcal{L}_{\rm SM}$.  Of course, $\mu_h^2=\lambda_h v_h^2$ and
$\lambda_h= m_h^2/2v_h^2$, where $v_h$ denotes the vacuum expectation
value (VEV) of the Higgs field at zero temperature, \emph{i.e.},
$\sqrt{2}\,\langle H \rangle_{T=0}=v_h$.  The scalar field $S$,
though, should have a vanishing VEV at zero temperature,
\textit{i.e.}\ $\langle S \rangle_{T=0}=0$, so as to maintain the
$\mathbb{Z}_3$ symmetry. Consequently, the quartic coupling $\lambda_s$ and the
portal coupling $\lambda_{hs}$ may be treated as free input
parameters, subject only to the requirements of perturbativity and
tree-level unitarity being preserved, namely $|\lambda_s|,\,
|\lambda_{hs}| \lesssim
4\pi$~\cite{Zhou:2020ojf,Chiang:2020yym,Athron_2018}.  Furthermore,
for the scalar potential to remain bounded from below, the quartic
couplings must satisfy~\cite{Kannike:2012pe}
\begin{equation}
\lambda_h>0,
\qquad
\lambda_s>0,
\qquad
\lambda_{hs}>-2\sqrt{\lambda_h\lambda_s}.
\end{equation}
Finally,  the bare mass
  parameter $\mu_s^2$ may be expressed in terms of the physical scalar
  mass as
\begin{equation}\label{eq:musq}
    \mu_s^2 = m_S^2 - \frac{1}{2}\lambda_{hs} v_h^2 .
\end{equation}
The cubic coupling $\mu_3$ would play a significant role in the
phase transition. It could very well be a complex quantity, thereby
engendering CP violation and the consequent mass splitting between the
scalar and pseudoscalar components of $S$. However, this is not
germane to the main arguments of the paper, and we would, henceforth,
treat $\mu_3$ to be real.

%% file: thermalHistory4.tex
\section{The scalar sector and thermal evolution}
\label{section3:pt}

With an additional scalar in play, the vacuum structure is more
complicated than within the SM. In particular, the quantum
corrections, and especially the temperature-dependent part may change
substantially.  Depending on the temperature, four distinct phases are
possible. These are identified by which of the two symmetries, the
electroweak and $\mathbb{Z}_3$ are broken or intact. At low
temperatures, we would, of course, want the former to be broken and
the latter preserved. And at very high temperatures, presumably both
are unbroken. Intriguingly, in between, there can exist a phase where
the electroweak symmetry is unbroken but the $\mathbb{Z}_3$ is not.

Since the particle masses and mixing structure depend on the vacuum
expectation values of the scalar fields, it is important to understand
the thermal evolution of the scalar potential.  To study this
evolution, we begin by discussing the finite-temperature effective
potential in terms of the background scalar fields, parametrizing these
as
\begin{equation}
H=
\begin{pmatrix}
0\\
h/\sqrt{2}
\end{pmatrix},
\qquad
S=\frac{1}{\sqrt{2}}(s+i\zeta),
\end{equation}
where $h$ denotes the neutral Higgs direction, while $s$ and $\zeta$
correspond respectively to the CP-even and CP-odd components of the
complex scalar field. The tree-level scalar potential is then
\begin{equation}
V_0(h,s)=
-\frac{1}{2}\mu_h^2 h^2
+\frac{1}{4}\lambda_h h^4
+\frac{1}{4}\lambda_{hs} h^2 s^2
+\frac{1}{2}\mu_s^2 s^2
+\frac{1}{3}\mu_3 s^3
+\frac{1}{4}\lambda_s s^4 \, ,
\end{equation}
and the finite-temperature evolution of this potential
determines the realization of the symmetry at different
temperatures.

The field-dependent scalar masses obtained from the tree-level scalar
potential are given by
\begin{equation}\label{eq:mass_fields}
  \barr{rcl}
&&  m_h^2(h,s)
=  \dis
-\mu_h^2
+3\lambda_h h^2
+\frac{\lambda_{hs}}{2}s^2,
\\[1.5ex]
&& m_{\chi_i}^2(h,s)
= \dis
-\mu_h^2
+ \lambda_h h^2
+\frac{\lambda_{hs}}{2}s^2,
\\[1.5ex]
&&m_s^2(h,s)
=\dis
\mu_s^2
+3\lambda_s s^2
+\frac{\lambda_{hs}}{2}h^2
+2\mu_3 s,
\\[1.5ex]
&&m_\zeta^2(h,s)
=\dis
\mu_s^2
+\lambda_s s^2
+\frac{\lambda_{hs}}{2}h^2
-2\mu_3 s,
\earr
\end{equation}
where $\chi_i$ are the Goldstone bosons related to the Higgs
doublet. In general, the CP-even mass matrix may contain
field-dependent off-diagonal terms. However, in the model scenario
under consideration, $H$ and $S$ never acquire VEVs simultaneously,
and, hence, such terms do not appear.

As for the rest of the particles, the expressions are as in the SM, with
$v_h$ being replaced by $h$ (to reflect field-dependent masses), namely
\begin{equation}
m_i^2(h)=\frac{1}{2}y_i^2h^2, \qquad
m_{W}^2(h)=\frac{1}{4}g^2h^2, \qquad
m_{Z}^2(h)=\frac{1}{4}(g^2+g^{\prime 2})h^2,
\end{equation}
where $y_i=\sqrt{2}m_i/v_h$ denotes the Yukawa coupling, while $g$ and
$g^\prime$ are the gauge couplings associated with the $SU(2)_L$ and
$U(1)_Y$ gauge groups of the SM. Whether it be for these or for the scalars,
the temperature corrections are straightforward. It should be borne in mind, though, that the photon too is rendered massive at
non-zero temperatures. Consequently, the Weinberg angle too receives
a temperature correction. 
The corresponding expressions are given in
App.\,\ref{app:thermal}.

At zero temperature, the field $S$ would necessarily have a vanishing
VEV. Consequently, the CP-even and CP-odd components of the complex
scalar field $S$ remain mass-degenerate, irrespective of the value of
the trilinear coupling $\mu_3$.  At finite temperatures, however, the
$\mu_3 s$ term induces a mass splitting (see
  Eq.\,\eqref{eq:mass_fields}), and this
  would turn out to play a crucial role in the realization of the
intermediate $\mathbb{Z}_3$-broken phase.

Within the $\overline{MS}$ scheme, the one-loop Coleman-Weinberg (CW)
correction at zero temperature is given by~\cite{Quiros:1999jp}
\begin{equation}
V^{\rm CW}_{\rm 1-loop}(h,s)
=
\pm
\frac{1}{64\pi^2}
\sum_i
n_i m_i^4(h,s)
\left[
\log\left(
\frac{m_i^2(h,s)}{\mu^2}
\right)
-C_i
\right],
\end{equation}
where the positive (negative) signs correspond to bosonic (fermionic)
fields, and $\mu$ denotes the renormalization scale. The multiplicity
factor $n_i$ is $12$ for SM quarks, $4$ for leptons, $2$ for Majorana
fermions, and $1$ for a real scalar or each polarization state of a
gauge boson. The constants $C_i$ are $3/2$ for scalars, fermions, and
longitudinal gauge-boson modes, and $1/2$ for transverse gauge-boson
modes.

The inclusion of radiative corrections through the CW potential shifts
the electroweak vacuum and modifies the scalar mass spectrum,
 and the tree-level minimization conditions are
no longer satisfied. To preserve the tree-level vacuum structure and
the physical scalar masses at zero temperature, appropriate
counterterms are introduced so that the renormalized effective
potential continues to satisfy the tree-level minimization
conditions~\cite{Carrington:1991hz,Quiros:1999jp}. The corresponding
counterterm potential is given by
\begin{eqnarray}\label{eq:ct}
V_{\rm ct}(h,s) &=& \frac{1}{2}\delta\mu_h^2 h^2
+\frac{1}{2}\delta\mu_s^2 s^2
+\frac{1}{4}\delta\lambda_h h^4 ,
\end{eqnarray}
where the counterterm coefficients $\delta\mu_h^2$, $\delta\mu_s^2$, and $\delta\lambda_h$ are fixed by imposing the zero-temperature renormalization conditions described in App.\,\ref{app:ct}.

The finite-temperature one-loop contribution is given by~\cite{Dolan:1973qd,Weinberg:1974hy}
\begin{equation}
V_T(h,s,T)
=
\frac{T^4}{2\pi^2}
\left[
\sum_B n_B
J_B\left(
\frac{m_B^2(h,s)}{T^2}
\right)
-
\sum_F n_F
J_F\left(
\frac{m_F^2(h,s)}{T^2}
\right)
\right],
\end{equation}
where $J_B$ and $J_F$ are the thermal bosonic and fermionic functions
discussed in App.\,\ref{app:thermal}.
The finite-temperature infrared divergences associated with
bosonic zero modes are treated through daisy resummation, ensuring the
perturbative consistency of the finite-temperature effective
potential~\cite{Espinosa:1995se,Carrington:1991hz,Arnold:1992rz}. Here
too, this is implemented along with replacing the field dependent
bosonic masses with their thermally corrected (Debye) masses in both
the zero and finite temperature contributions~\cite{Parwani:1991gq},
leading to
\begin{equation}
m_i^2(h,s,T)=m_i^2(h,s)+\Pi_iT^2,
\end{equation}
where $\Pi_i$ represent the thermal self-energy coefficients,
determined from the zero momentum limit of the corresponding two-point
function\footnote{The transverse gauge modes do not acquire a static
thermal mass at leading order, $\Pi_T(0,\mathbf{k}\to0)=0$, since
static magnetic fields are not perturbatively screened; magnetic
screening arises only nonperturbatively at the scale
$\mathcal{O}(g^2T)$~\cite{Altherr:1993tn,Weldon:1982aq,Gross:1980br}.}. The
explicit expressions are given in App.\,\ref{app:thermal}.  The
resulting one-loop finite-temperature effective potential is, therefore,
given by
\begin{equation}\label{eq:effpot}
V_{\rm eff}(h,s,T)
=
V_0(h,s)
+
V^{\rm CW}_{\rm 1-loop}(h,s,T)
+
V_T(h,s,T)
+
V_{\rm ct}(h,s)\ .
\end{equation}
It should be noted that perturbative analyses of finite-temperature
phase transitions are subject to two primary theoretical
uncertainties: the choice of the renormalization scale and the gauge
dependence of the effective potential. Within the present
regularization framework, $\mu$ introduces an intrinsic uncertainty in
the determination of the critical temperature and related
observables~\cite{Chiang:2018gsn,Athron:2022jyi,Croon:2020cgk,Gould:2021oba}.
Throughout this work, we fix the renormalization scale to the top
quark mass\footnote{For a critical temperature around the electroweak
scale, changing the renormalization scale from $\mu=m_t/2$ to
$\mu=2m_t$ induces only moderate changes in the PT parameters,
typically less than $\mathcal{O}(10\%)$~\cite{Athron:2022jyi}.},
$\mu=m_t$. The explicit renormalization scale dependence of the
one-loop CW potential can be reduced by employing an RGE-improved
treatment~\cite{Andreassen:2014eha,Andreassen:2014gha}. Furthermore,
our numerical analysis is performed in the Landau gauge using the
conventional one-loop finite-temperature effective potential. Since
the EWPT in the present model is driven primarily by the tree-level
Higgs-portal interaction, the gauge dependence arising from loop-level
corrections is expected to be subleading for the qualitative features
of our analysis. Although this framework is widely employed in
phenomenological studies, residual gauge, infrared, and
resummation-scheme uncertainties can affect precision predictions of
SFOPT~\cite{Croon:2020cgk,Balui:2025yvd}. A quantitatively robust
determination would require dimensional reduction to a
three-dimensional effective field theory, supplemented by higher-order
thermal matching and, where necessary, non-perturbative lattice
simulations~\cite{Chala:2024xll,Chala:2025oul,Bernardo:2026whs}.

Using this formalism, we now discuss the various phase transitions governed by the spontaneous breaking of the $\mathbb{Z}_3$ and electroweak symmetries for suitable regions of parameter space.

\subsection{High-temperature symmetric phase}
At sufficiently high temperatures, the thermally corrected scalar mass terms
dominate the effective potential, rendering the effective mass-squared
terms of both the scalar fields positive. As a result, the
origin of the field space, $(h,s)=(0,0)$, remains the global minimum,
and both the electroweak and $\mathbb{Z}_3$ symmetries stay unbroken.

In this phase, all particles are relativistic and remain in thermal
equilibrium with the primordial plasma. The equilibrium number density
scales as $n\sim T^3$, while for renormalizable interactions the
thermally averaged annihilation cross section typically scales as
$\langle \sigma v \rangle \sim 1/T^2$. Consequently, the interaction
rate behaves as $\Gamma \sim n\langle \sigma v \rangle \sim T$,
whereas the Hubble expansion rate during the radiation-dominated era
scales as $H \sim T^2/M_{\rm Pl}$. Therefore,  $\Gamma/H\sim M_{\rm Pl}/T$ which is much larger than unity throughout the temperature range relevant to our analysis ($T\ll M_{\rm Pl}$). Hence, all particles remain in thermal equilibrium. 

In this symmetric phase, the fermionic sector consists of the Dirac
fermion $\chi$ and the right-handed Majorana fermion $N_R$, while the
scalar sector contains the complex scalar $S$. 

\subsection{Two-step phase transition and intermediate $\mathbb{Z}_3$-broken phase}
As the Universe cools to a certain value $T_s$ (above the critical
temperature $T_c$), the thermal contributions gradually diminish, and
the curvature of the effective potential along the $S$ direction may
become negative, thereby leading to an intermediate phase in which the
dark scalar acquires a non-zero VEV (breaking the $\mathbb{Z}_3$
symmetry) while the electroweak symmetry remains preserved, namely
\begin{equation}
\langle H\rangle = 0,
\qquad
\sqrt{2}\,\langle S\rangle = v_s(T) \neq 0.
\end{equation}
The appearance and stability of this intermediate vacuum are governed by the thermal evolution of the finite-temperature effective potential. For an analytic analysis of the vacuum structure, the tree-level potential together with the high-temperature expansion of the finite-temperature corrections, valid for
$m^2(h,s,T)/T^2\ll1$ is employed~\cite{Ghorbani:2020xqv}.

As the temperature further decreases to the critical temperature
$T_c$, another local minimum emerges at $\bigl(v(T_c),0\bigr)$,
corresponding to the EW broken phase, and becomes degenerate with the
symmetric phase minimum $\bigl(0, v_s(T_c)\bigr)$. With further
cooling, this electroweak breaking minimum becomes the global minimum
of the potential, making it energetically favored over the electroweak
symmetric minimum at $\langle S\rangle\neq 0$. If a potential barrier
exists between the electroweak symmetric minimum at $\langle
S\rangle\neq 0$ and the electroweak breaking minimum at $\langle
S\rangle=0$, the EWPT proceeds as a first-order transition. In this
scenario, the potential barrier is generated by a sufficiently large
Higgs-portal coupling $\lambda_{hs}$, with its effect depending on the
mass of the scalar field.

The critical temperature $T_c$ is determined from the
finite-temperature effective potential by imposing two stationarity
conditions and one vacuum-degeneracy condition. These three conditions
ensure that the two phases correspond to local minima of the potential
with equal free energy at $T=T_c$, and are given by
\begin{eqnarray}
\left.\frac{\partial V_{\rm eff}(h,0,T_c)}{\partial h}\right|_{h=v(T_c)}
&=& 0,
\qquad
\left.\frac{\partial V_{\rm eff}(0,s,T_c)}{\partial s}\right|_{s=v_s(T_c)}=0,
\\[3mm]
V_{\rm eff}(v(T_c),0,T_c) &=& V_{\rm eff}(0,v_s(T_c),T_c).
\end{eqnarray} 
The dimensionless ratios 
\begin{equation}
\xi_h=\frac{v(T_c)}{T_c}, \qquad \xi_s=\frac{v_s(T_c)}{T_c},
\end{equation}
characterize the strength of the phase transition (PT) along the
  Higgs and singlet directions with a strongly first-order EWPT
requiring the order parameter, $\xi_h\gtrsim1$.  Although the two
vacua are energetically degenerate at $T_c$, the Universe remains
temporarily trapped in the false vacuum due to the presence of a
potential barrier separating the two phases. For $T<T_c$, the
transition proceeds via quantum tunneling from the metastable
electroweak symmetric phase to the electroweak breaking phase through
the nucleation of critical bubbles. The nucleation temperature $T_n$
is determined by the condition that the bubble nucleation probability
becomes of order unity within a Hubble volume, and hence satisfies
$T_n<T_c$ for a first-order phase transition, as discussed later in
the GW section.

\begin{table}[ht]
\centering
\begin{tabular}{|c|c|c|c|c|c|c|c|}
\hline
BPs &
$m_S$ [GeV] &
$\lambda_s$ &
$\lambda_{hs}$ &
$\mu_3$ [GeV] &
$m_\chi$ [GeV] &
$y_2$ &
$\Omega_{\rm DM} h^2$ \\
\hline

BP1 &
\multirow{3}{*}{100.0} &
1.50 &
1.10 &
-10.0 &
40.0 &
0.315 &
0.12 \\
\cline{1-1}\cline{3-8}

BP2 &
&
1.45 &
1.10 &
-10.0 &
80.0 &
0.279 &
0.12 \\
\cline{1-1}\cline{3-8}

BP3 &
&
2.0 &
1.20 &
-10.0 &
95.0 &
0.252 &
0.12 \\
\hline

BP4 &
\multirow{3}{*}{150.0} &
2.20 &
1.57 &
-10.0 &
40.0 &
0.454 &
0.12 \\
\cline{1-1}\cline{3-8}

BP5 &
&
2.28 &
1.60 &
-10.0 &
80.0 &
0.368 &
0.12 \\
\cline{1-1}\cline{3-8}

BP6 &
&
2.25 &
1.59 &
-10.0 &
120.0 &
0.349 &
0.12 \\
\hline

\end{tabular}
\caption{Benchmark points satisfying the observed dark matter relic density. For all benchmark points, we fix $y_1 = 0.01$ and $m_R = 10\,\mathrm{GeV}$.}
\label{tab:FOPT}
\end{table}

\begin{figure}
    \centering
    \includegraphics[width=0.48\linewidth]{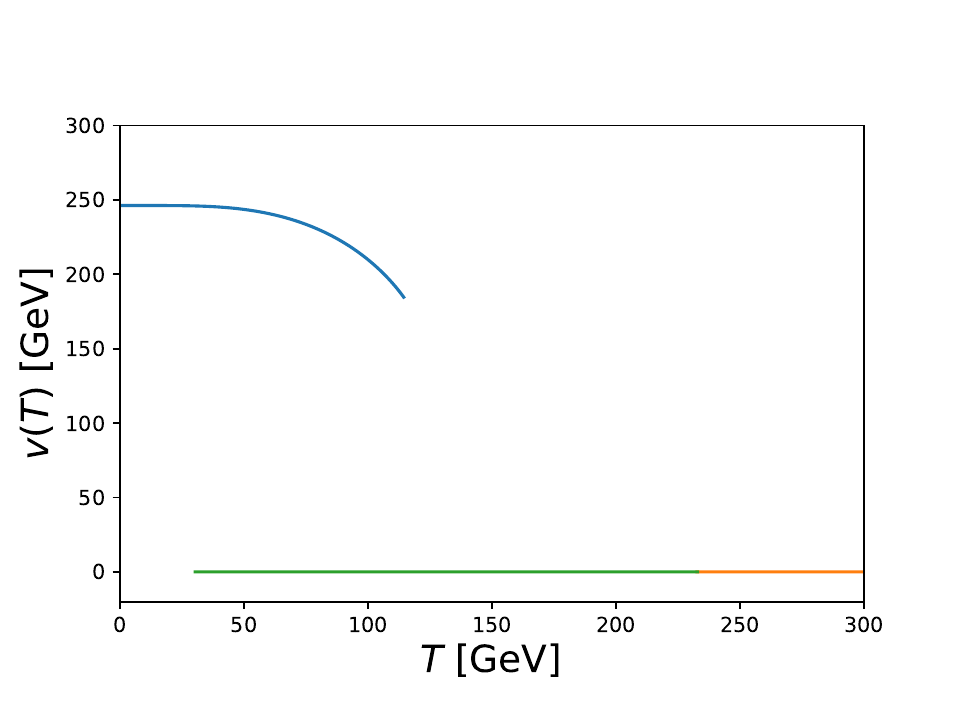}
    \includegraphics[width=0.48\linewidth]{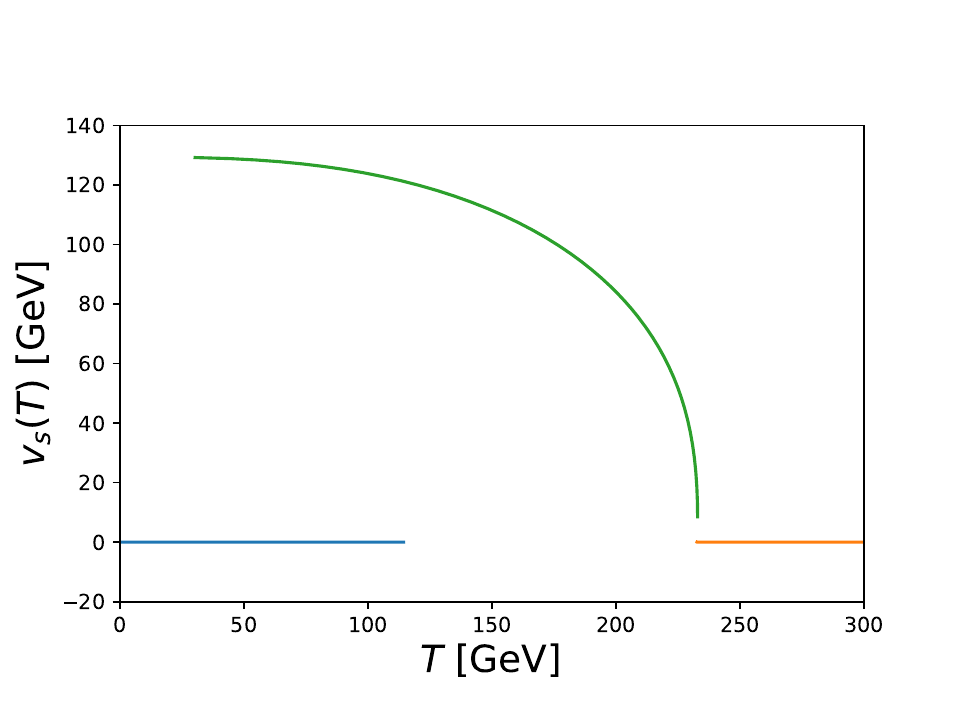}
    \caption{Temperature evolution of the VEVs for BP2. The left (right) panel shows the Higgs (singlet) VEV, with colors denoting the different phases of the phase transition.}
    \label{fig:bp2}
\end{figure}

The thermal evolution of the vacuum structure and the corresponding PT
parameters, including the critical temperature and VEVs, are computed
numerically using the publicly available code
\texttt{CosmoTransitions}~\cite{Wainwright:2011kj}. The resulting PT
parameters and order parameters for all BPs listed in
Tab.\,\ref{tab:FOPT} are presented in
Tab.\,\ref{tab:FOPT1}. As
an illustrative example, Fig.\,\ref{fig:bp2} shows the temperature
dependence of the Higgs and singlet VEVs for BP2. The evolution
exhibits an intermediate singlet phase at finite temperature, followed
by the transition to the electroweak vacuum as the temperature
decreases.

The intermediate $\mathbb{Z}_3$ broken vacuum significantly alters the scalar as well as the fermionic particle spectrum.

\begin{table}[ht]
\centering
\small
\setlength{\tabcolsep}{3.5pt}
\renewcommand{\arraystretch}{1.0}
\begin{tabular}{ccccccc}
\toprule
\textbf{PT param.} & \textbf{BP1} & \textbf{BP2} & \textbf{BP3} & \textbf{BP4} & \textbf{BP5} & \textbf{BP6} \\
\midrule
$(v,v_s)|_{T_c}^{\rm high}$ & $(0,123)$ & $(0,126)$  & $(0,112)$ & $(0,99)$  & $(0,100)$  & $(0,100)$  \\
$(v,v_s)|_{T_c}^{\rm low}$  & $(222,0)$& $(225,0)$ & $(217,0)$ & $(201,0)$& $(205,0)$ &  $(204,0)$\\
$T_c$  & 90 & 86& 94& 106 & 104 & 104\\
$(\xi_h,\xi_s)$ & $(2.47,1.37)$ & $(2.62,1.47)$ & $(2.31,1.19)$& $(1.90,0.93)$& $(1.97,0.96)$ & $(1.96,0.96)$ \\
$(v,v_s)|_{T_n}^{\rm high}$ &$(0,126)$ & $(0,129)$ & $(0,114)$ & $(0,104)$ &  $(0,105)$& $(0,105)$ \\
$(v,v_s)|_{T_n}^{\rm low}$  & $(242,0)$ & $(245,0)$ & $(232,0)$ & $(233,0)$ & $(237,0)$ & $(236,0)$ \\
$T_n$ & 57 & 40& 77 & 77& 70& 72 \\
$\alpha_n$ & 0.16 & 0.45 & 0.07 & 0.08 & 0.11 &0.10 \\
$\beta/H_n$ & 412& 286 & 1293& 876 & 394 & 260\\
\bottomrule
\end{tabular}
\caption{Phase transition parameters for the different benchmark points. VEVs and temperature are in units of GeV.}
\label{tab:FOPT1}
\end{table}

\subsubsection{Particle content in the $\mathbb{Z}_3$ broken phase}
The two degrees of freedom of the scalar acquire different masses in
the $\mathbb{Z}_3$-broken phase due to the cubic scalar interaction
proportional to $\mu_3$. The Dirac fermion $\chi$ splits into two
Majorana fermions which subsequently mix with the right-handed
Majorana fermion $N_R$.

Defining the CP-even (and CP-odd) Majorana field $\chi_1$ ($\chi_2$)
\begin{equation}
\chi_1=
\frac{1}{\sqrt{2}}(\chi+\chi^c),
\qquad
\chi_2=
\frac{-i}{\sqrt{2}}(\chi-\chi^c),
\qquad
N=N_R+N_R^c,
\end{equation}
the Lagrangian part containing the mass terms in the basis
$(\chi_1,\chi_2,N)$ takes the form\footnote{Note that the operator
$\bar{\chi}_2 i\gamma_5 N$ is hermitian.}
\begin{eqnarray}
    \mathcal{L}_M = - \frac{1}{2}\overline{\Psi} M(T) \Psi - \frac{1}{2}\big(m_\chi-\sqrt{2}y_1v_s(T)\big)\overline{\chi}_2\chi_2 -\frac{ (y_3-y_2)v_s(T)}{2} \overline{\chi}_2 i\gamma_5 N,
\end{eqnarray}
where the mass matrix in the basis $\Psi=(\chi_1,N)^T$ takes the form

\begin{equation}
M(T)=
\begin{pmatrix}
m_\chi+\sqrt{2}y_1v_s(T)  & \qquad \qquad & 
\dfrac{(y_2+y_3)v_s(T)}{2}
\\[3mm]
\dfrac{(y_2+y_3)v_s(T)}{2} & &
m_R
\end{pmatrix}.
\end{equation}

The complete mass matrix is, of course a $3 \times 3$ one and it
  is straightforward to diagonalize it and obtain the mass eigenstates
  $P_i$.  It is instructive, however, to consider the limit $y_2 \to
  y_3$ when $\chi_2$ decouples from the both $\chi_1$ and $N$. In this
  limit, we have $P_3 = \chi_2$ with a temperature dependent mass
  $m_{\chi_2}(T) = m_\chi-\sqrt{2}y_1v_s(T)$. As for the other two
  eigenstates, these can, in this limit, be expressed through
\begin{equation}
\begin{pmatrix}
\chi_1\\
N
\end{pmatrix}
=
U(T)
\begin{pmatrix}
P_1\\
P_2
\end{pmatrix},
\qquad
U^T(T)\mathcal{M}(T)U(T)
=
{\rm diag}(m_1(T),m_2(T)),
\end{equation}
where $U(T)$ is a temperature dependent real $2\times 2$ orthogonal matrix given by 
\begin{eqnarray}
    U(T)= \begin{pmatrix}
        \cos\theta(T) & \sin\theta(T)\\[0.5mm]
       - \sin\theta(T) & \cos\theta(T) 
    \end{pmatrix}, \,\, \theta(T) = \frac{1}{2}\arctan\left(\frac{(y_2+y_3)v_s(T)}{m_R-(m_\chi+\sqrt{2}y_1v_s(T))}\right).
\end{eqnarray}
Note that $N=\displaystyle\sum_{i=1}^2 U_{2i}P_i= -  \sin\theta(T) P_1 +  \cos\theta(T) P_2$ .

The Yukawa interactions of the scalar and pseudoscalar with fermions,
in the $(\chi_1,\chi_2, N)$ basis, are easily seen to be
\begin{equation}
\barr{rcl}
\mathcal{L}_{s} &= & \dis
-\frac{y_1}{\sqrt{2}}\bar{\chi}_1\chi_1 s
+ \frac{y_1}{\sqrt{2}}\bar{\chi}_2\chi_2 s
-\frac{ (y_3+y_2)}{2} \bar{\chi}_1 N s
- \frac{ (y_3-y_2)}{2} \bar{\chi}_2 i\gamma_5 N s,
\\[2.5ex]
\mathcal{L}_{\zeta} & = & \dis
\sqrt{2}\,y_1\bar{\chi}_1\chi_2 \zeta
-\frac{ (y_3+y_2)}{2} \bar{\chi}_2 N \zeta
+ \frac{ (y_3-y_2)}{2} \bar{\chi}_1 i\gamma_5 N \zeta . 
\earr
\end{equation}
The same can be expressed in the fermion mass-basis in a straightforward
  manner and are presented in App.\,\ref{app:couplings}.
 For the
parameter space yielding the correct DM abundance, we find 
that all
three states remain in thermal equilibrium through efficient
annihilation, conversion, decay, and scattering processes such as
\begin{equation}
P_iP_j \leftrightarrow P_kP_l,\qquad P_iP_j \leftrightarrow XX,
\qquad
X=s,\zeta,H,
\end{equation}
and 
\begin{equation}
    P_{1,2}\to L+H ,\qquad \chi_2 \leftrightarrow P_{1,2}+\zeta.
\end{equation}
 It is the sizeable Higgs portal interaction that keeps
 the dark sector in equilibrium with the SM plasma. Consequently, the
 dark and visible sectors share a common temperature throughout the
 $\mathbb{Z}_3$-broken phase.

The intermediate $\mathbb{Z}_3$-broken phase exists only over a finite
temperature interval and terminates well before the eventual DM
freeze-out. Thus, the relic density doesn't directly depend on the
stability of intermediate mass eigenstates. Rather, it is sufficient
that the dark sector remains thermalized throughout this phase. For
the parameter space which simultaneously yields the observed relic
abundance and a SFOEWPT, all dark states are relativistic or only
mildly non-relativistic throughout the $\slashed{\mathbb{Z}}_3$
phase. Their number densities are therefore not significantly
Boltzmann suppressed.

\subsection{EWPT }
At lower temperatures, the Higgs field acquires a non-zero
VEV\footnote{Note that, at $T=0$, $v(0)=v_h\approx 246$ GeV.} while
the singlet VEV vanishes,
\begin{equation}
\sqrt{2}\,\langle H\rangle=v(T),
\qquad
\langle S\rangle=0.
\end{equation}
Consequently, the electroweak symmetry is spontaneously broken,
whereas the $\mathbb{Z}_3$ symmetry is restored. The fermionic
spectrum therefore reorganizes back into the original
(zero-temperature) interaction basis consisting of the Dirac fermion
$\chi$ and the right-handed Majorana fermion $N_R$.
The thermal evolution of the vacuum thus follows the sequence
\begin{eqnarray}
    (\rm EW,~\mathbb{Z}_3) \to (\rm EW,~\slashed{\mathbb{Z}}_3) \to (\slashed{\rm EW},~\mathbb{Z}_3).\nonumber
\end{eqnarray}
As mentioned earlier, the PT parameters resulting from the SFOEWPT are
summarized in Tab.\,\ref{tab:FOPT1}.  For larger scalar masses, a
stronger Higgs portal coupling $\lambda_{hs}$ is required to realize
the transition\footnote{This behavior readily follows from the
relation $\mu_S^2(T) \approx m_S^2-\frac{1}{2}\lambda_{hs}v_h^2 +
\Pi_{ss} T^2$ (valid only for the high-temperature approximation
without CW contributions). At intermediate temperatures where
$\mathbb{Z}_3$ is spontaneously broken, $\mu_S^2(T) < 0$. Since
$\Pi_{ss} T^2$ is not too large for temperatures of interest, as $m_S$
increases, a correspondingly larger $\lambda_{hs}$ is required to
drive the effective mass-squared negative. As for the Yukawa coupling
contributions to $\mu_S^2(T)$, these appear only through the one-loop
thermal correction contained in $\Pi_{ss}$ and are further suppressed
by the small values of the Yukawa coupling, as mandated by the
direct-detection constraints. Consequently, its effect on the
finite-temperature effective potential is subdominant compared to the
tree-level contribution from the Higgs--portal coupling
$\lambda_{hs}$, which therefore primarily controls the realization of
a SFOEWPT.}. Consequently, excessively large scalar masses demand
large values\footnote{For illustration, the variation of the
Higgs-portal coupling $\lambda_{hs}$ with the singlet scalar mass
$m_S$ subject to a strongly first-order EWPT can be found in
Refs.~\cite{Vaskonen:2016yiu,Cline:2013gha,Chaudhuri:2022sis}.} of $\lambda_{hs}$, which may eventually violate
perturbative unitarity.

For BP1--BP3, where the singlet scalar mass lies close to the Higgs
boson mass, the PT order parameters, $\xi_h$ and $\xi_s$, are large in
both field directions, indicating a SFOPT in both the Higgs and
singlet fields. In contrast, for BP4--BP6, corresponding to larger
singlet scalar masses, the order parameter is appreciable only along
the Higgs direction, while it is significantly suppressed along the
singlet direction. This behavior can be understood as a consequence of
the decoupling of the heavier singlet from the Higgs sector. Although
a larger portal coupling is required as the singlet scalar mass
increases, the heavy singlet gradually decouples from the Higgs sector
due to considerable Boltzmann suppression. As a result, its role in
driving the EWPT is diminished, leading to a weaker first-order
transition characterized by smaller values for the order
parameters. The cubic coupling $\mu_3$ controls the tree-level barrier
in the singlet direction, while the Higgs-portal coupling
$\lambda_{hs}$ controls the interplay between the singlet and Higgs
directions, with their combined effect determining the strength of the
first-order EWPT.

\subsubsection{DM in the restored $\mathbb{Z}_3$ phase}
We now discuss the corresponding DM phenomenology in the restored
$\mathbb{Z}_3$ phase.  After the restoration of the $\mathbb{Z}_3$
symmetry, the DM phenomenology is governed by the Dirac fermion $\chi$
and the complex scalar $S$. It might seem that, with both $\chi$ and
$S$ being charged under $\mathbb{Z}_3$, the lighter of the two could
be the DM candidate with the heavier entity decaying into the DM-$\nu$
pair. However, note that the $\lambda_{hs}$ term is entirely oblivious
to the $\mathbb{Z}_3$ charge of $S$, and, post EWSB, mediates the
process $S +f \to S +f$, for any SM fermion $f$ through the
($t$-channel) Higgs portal. Thus, the relatively large $\lambda_{hs}$
that is required to drive a SFOEWPT would run counter to the
constraints on direct detection rates from, say the LZ
experiment~\cite{LZ:2024zvo}.  Given this, we would consistently work
in the regime with $m_\chi < m_S$, rendering the $\chi$ the dominant
DM constituent.

\begin{figure}[!h]
    \centering
    \includegraphics[width=0.49\linewidth]{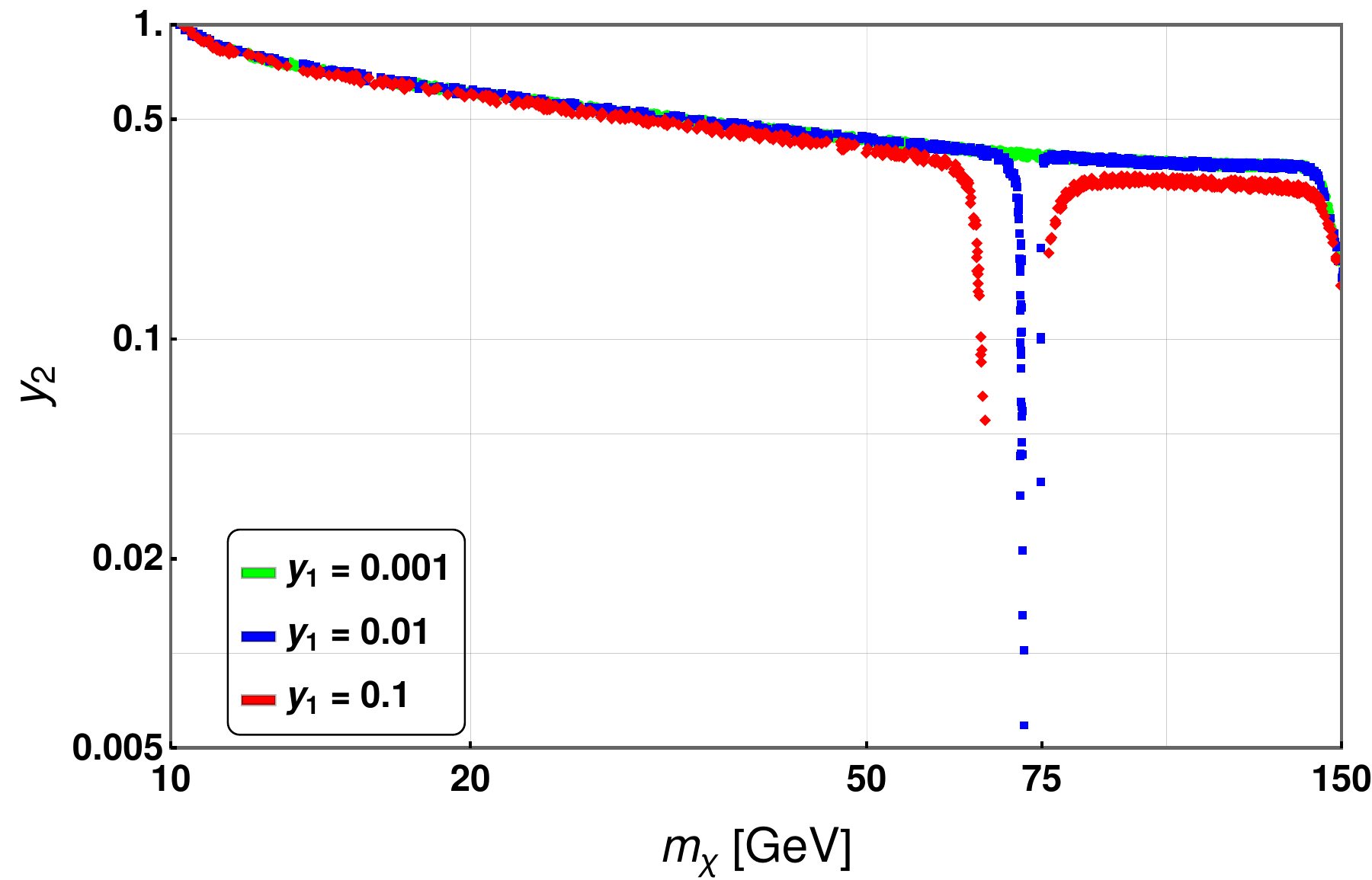}
    \includegraphics[width=0.49\linewidth]{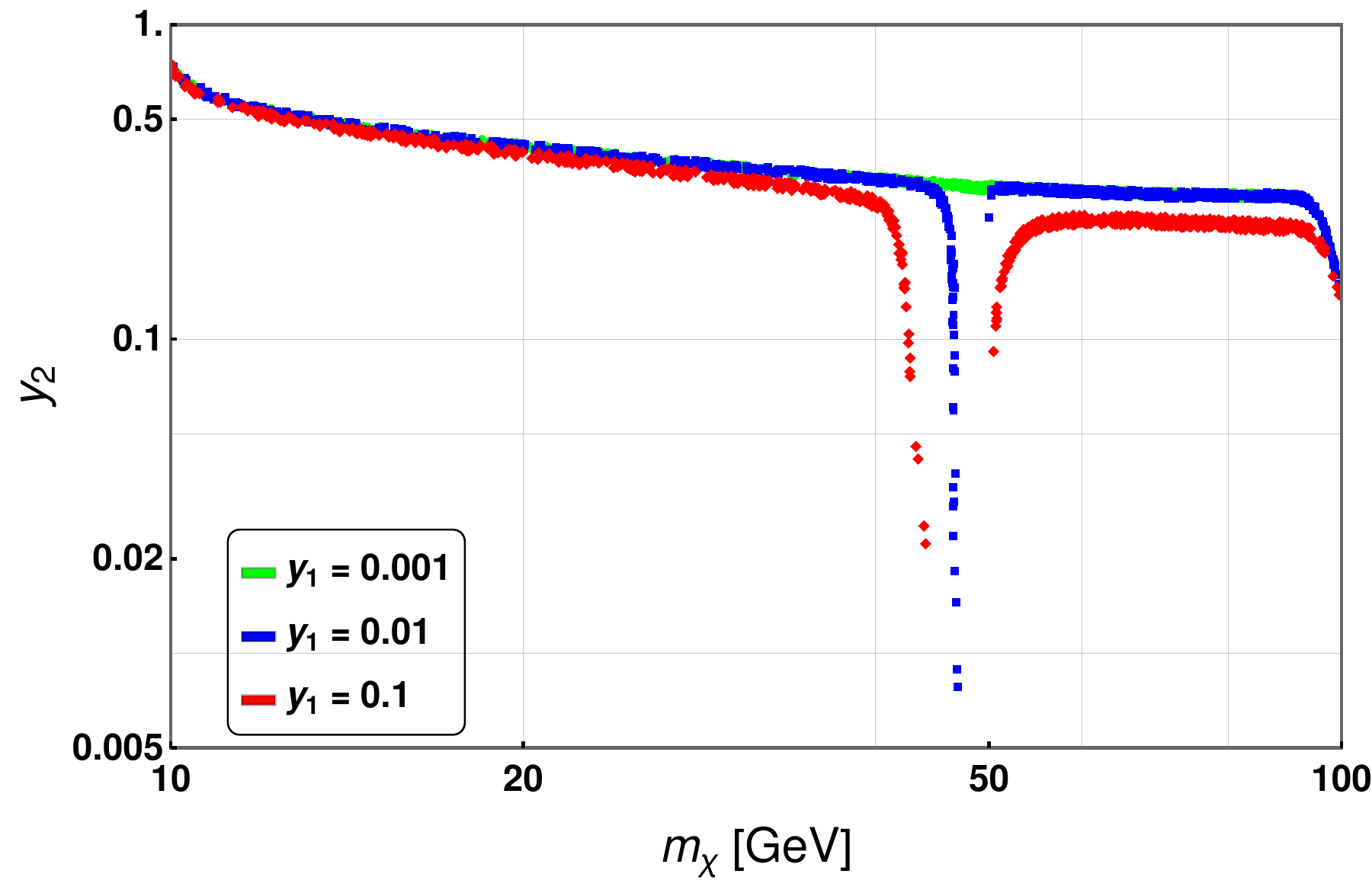}
    \caption{Scatter plot in the $y_2$--$m_\chi$ plane satisfying the
      relic abundance constraint $0.1 \leq \Omega_{\mathrm{DM}} h^2
      \leq 0.121$, for $\mu_3=-10~\mathrm{GeV}$ and
      $m_R=10~\mathrm{GeV}$. Red, blue, and green points correspond to
      $y_1=0.1$, $y_1=0.01$, and $y_1=0.001$, respectively. The left
      panel corresponds to $m_S=150~\mathrm{GeV}$ and
      $\lambda_{hs}=1.5$ while the right one is for
      $m_S=100~\mathrm{GeV}$ and $\lambda_{hs}=0.8$.}
    \label{fig:relicN}
\end{figure}

\begin{figure}
    \centering
    \includegraphics[width=0.49\linewidth]{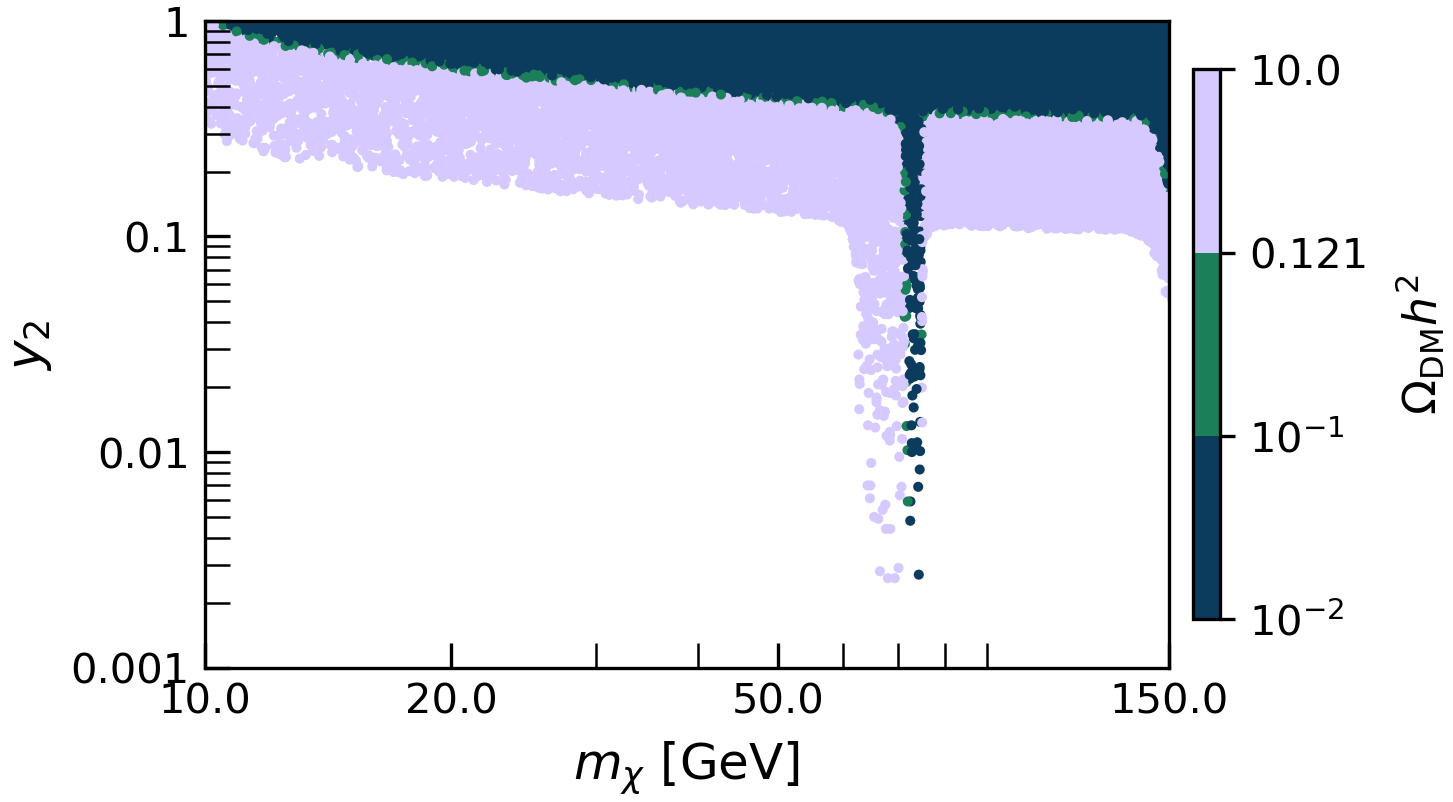}
    \includegraphics[width=0.49\linewidth]{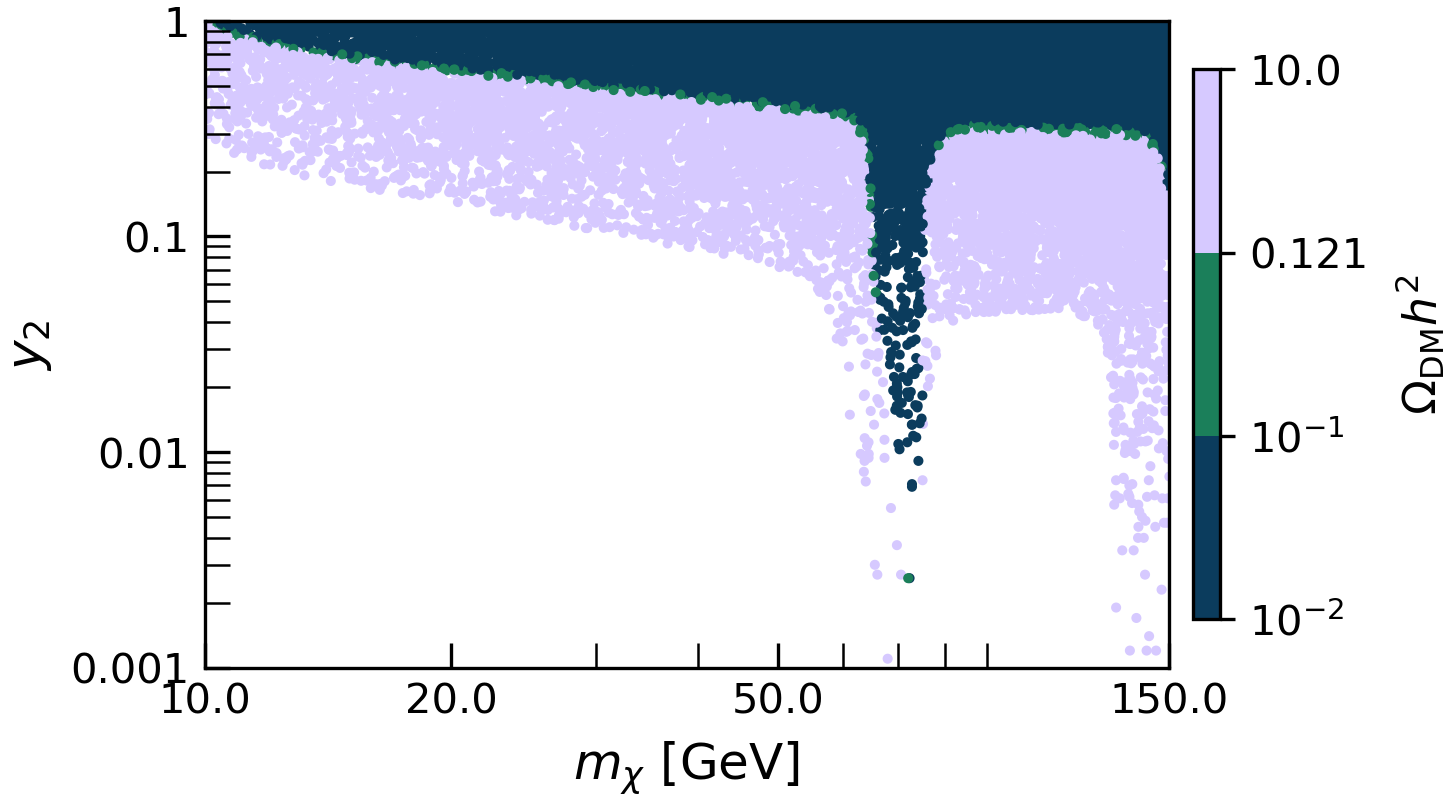}
     \caption{Scatter plot in the $y_2$--$m_\chi$ plane for $\lambda_{hs}=1.5$, $\mu_3=-10~\mathrm{GeV}$, $m_S=150~\mathrm{GeV}$, and $m_R=10~\mathrm{GeV}$. Different colors indicate different ranges of the relic density. The left panel corresponds to $y_1=0.01$, while the right panel corresponds to $y_1=0.1$.}
    \label{fig:relicN_band}
\end{figure}

\begin{figure}
    \centering
    \includegraphics[width=0.49\linewidth]{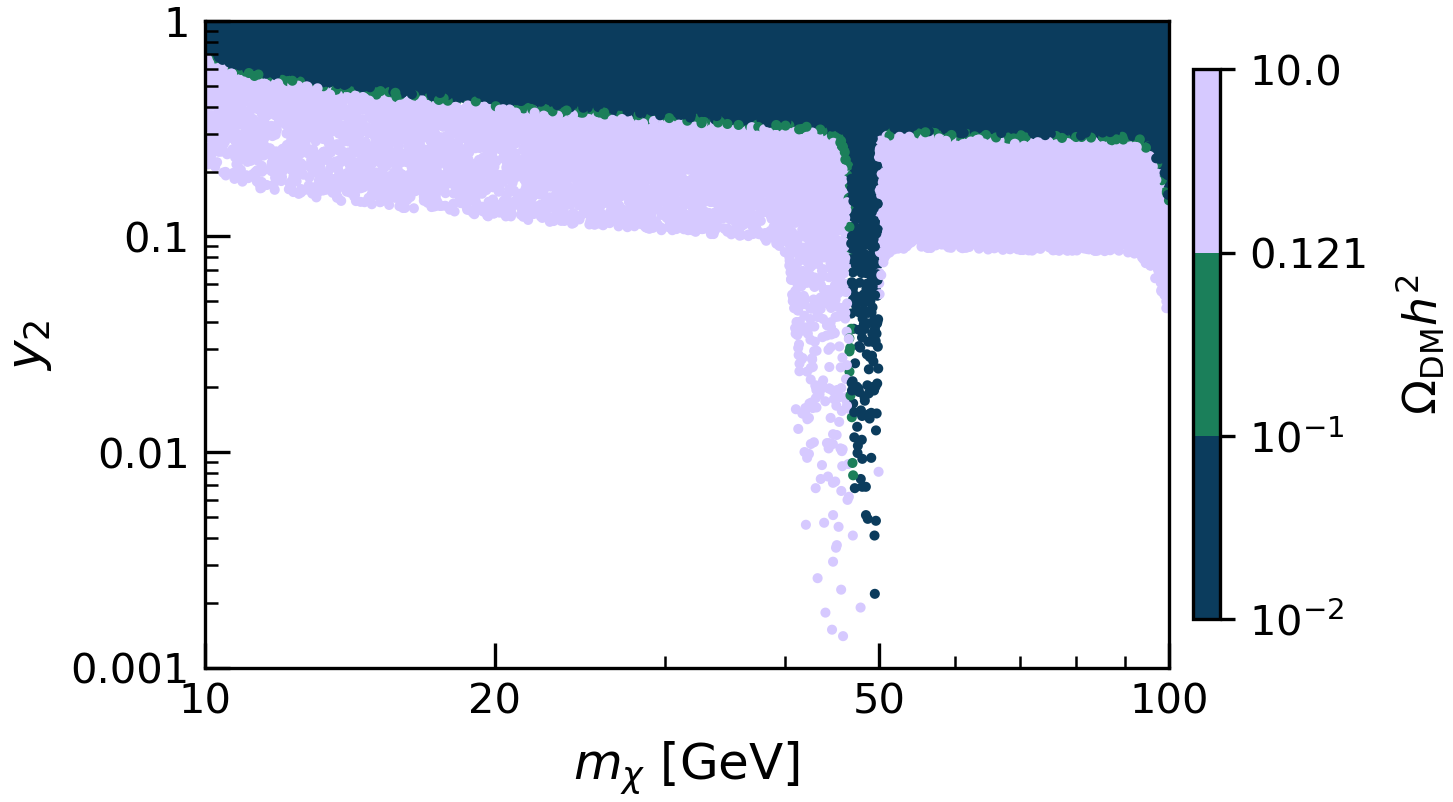}
    \includegraphics[width=0.49\linewidth]{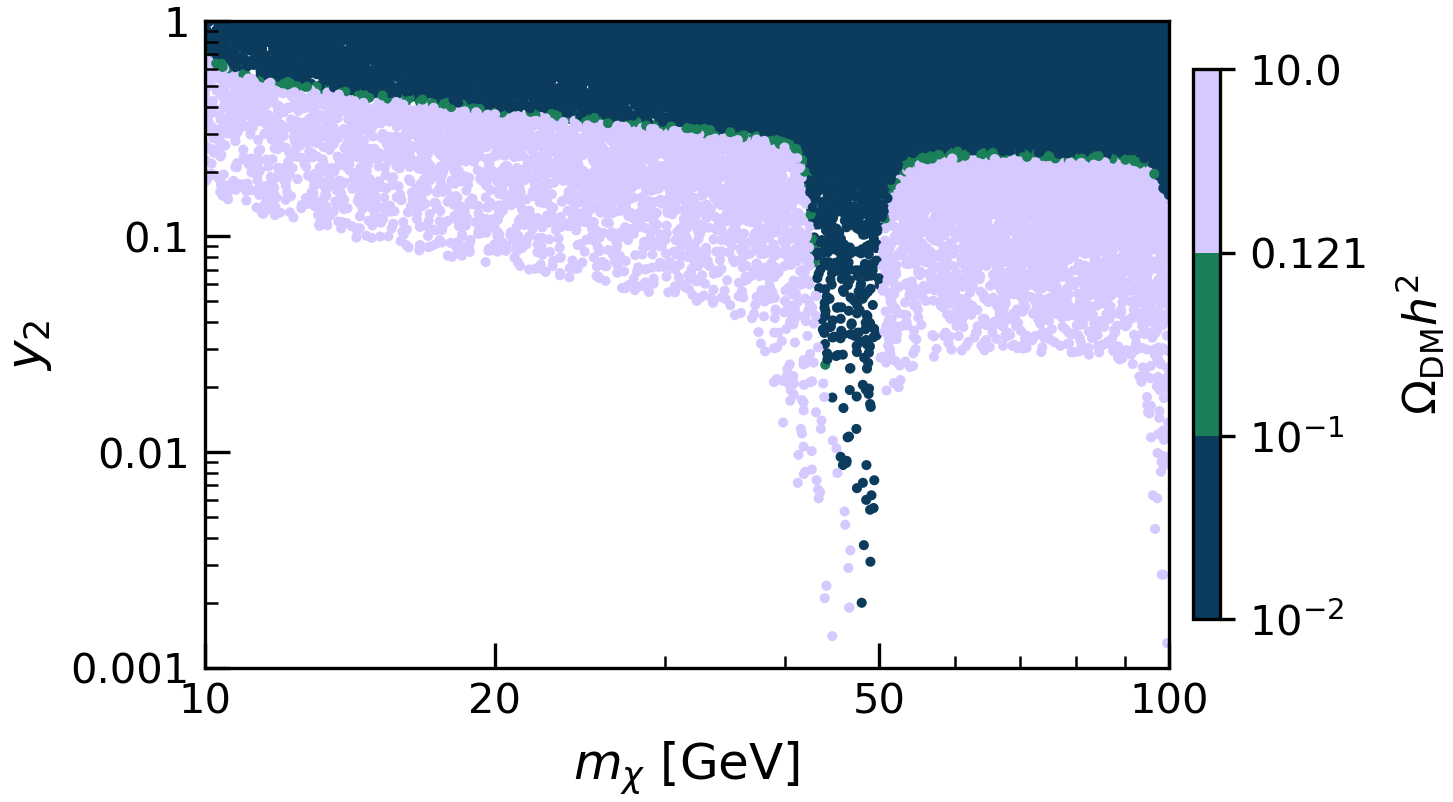}
    \caption{The same as Fig.\,\ref{fig:relicN_band}, shown for $m_S = 100\,\mathrm{GeV}$ and $\lambda_{hs}=0.8$.}
    \label{fig:relicN_band100}
\end{figure}

The dominant processes contributing to the fermionic DM relic abundance include annihilation, semi-annihilation, and dark-sector conversion channels:
\begin{eqnarray}
\label{eqN}\chi\chi \rightarrow \chi^c N,&
\qquad& \bar{\chi}{\chi} \rightarrow NN,
\\ \label{eqS}
\chi\chi \rightarrow SS,
&\qquad&
\chi\chi \rightarrow HS^\dagger,
\qquad
\bar{\chi}\chi \rightarrow SS^\dagger.
\end{eqnarray}
In particular, the $S\bar{\chi}N_R$ and cubic scalar interaction term
leads to the semi-annihilation channels.  The relic abundance and
thermal freeze-out dynamics are computed numerically using {\tt
  micrOMEGAs}~\cite{Belanger:2018ccd}. The parameter regions
reproducing the observed DM relic abundance consistent with the latest
Planck observations~\cite{Planck:2018vyg} are shown in
Figs.~\ref{fig:relicN}-~\ref{fig:relicN_band100}.

To interpret the parameter space depicted in figures, we focus on the
representative benchmark choice $m_R=10~{\rm GeV}$ adopted throughout
this analysis. A relatively light RHN opens efficient annihilation and
semi-annihilation channels involving $N$, thereby enlarging the
parameter space compatible with the observed relic abundance. For this
relatively light RHN, the processes in Eq.~\eqref{eqN} generally
dominate over those in Eq.~\eqref{eqS}, except near kinematic
thresholds where scalar-sector processes may become important.

If $m_\chi > m_S$, the fermion $\chi$ decays into the lighter scalar
state, making the scalar the dominant DM component. However, as
mentioned earlier, the scalar DM scenario is strongly constrained by
the combined requirements of a SFOPT, relic abundance, and direct
detection limits and we would not discuss it any further.

Concentrating on $m_\chi < m_S$, the processes listed in
  Eq.~(\ref{eqN}) contribute most to the relic density
  evolution. Fig.\,\ref{fig:relicN} illustrates the correlation in the
  $y_2$--$m_\chi$ plane, (for three representative values of $y_1$)
  that is needed to achieve the right relic density. It is easy to see
  that the sensitivity to $y_1$ is low while that to $y_2$ is
  considerable. This is understandable as the latter controls the
  interaction strength between the dark fermion and the RHN. This is
  better understood from Figs.\,\ref{fig:relicN_band} and
  \ref{fig:relicN_band100} which show scatter plots in the
  $y_2-m_\chi$ plane with the points color-coded according to the
  levels of the relic density. The parameter space scan is performed over the region $m_\chi\leq m_R \leq m_S$.  For $m_\chi < m_R$,
reproducing the observed relic abundance typically requires substantially larger Yukawa couplings $y_2$ as annihilation and semi-annihilation channels become phase-space suppressed. The
condition $m_\chi\leq m_S$ ensures that the dark fermion remains the lightest
$\mathbb{Z}_3$-charged state and hence the DM candidate.

There are two prominent features in these plots. The dip around
$m_\chi \simeq m_S/2$ originates from the resonant enhancement of the
semi-annihilation process $\chi\chi \to \chi N$ mediated by an
$s$-channel scalar. A second dip appears near $m_\chi \simeq m_S$ due
to the opening of the $\chi\chi \to SS$ channel. In particular, for $m_\chi
  \simeq m_S$, the processes listed in Eq.~\eqref{eqS} become
  comparable to, and can even dominate over, those in
  Eq.~\eqref{eqN}. The process $\chi\chi \to SS$ receives
contributions from the $t$- and $u$-channel exchange of $\chi$, as
well as an $s$-channel scalar exchange mediated by the cubic
interaction $\mu_3 S^3$. In this region, the velocity of final-state
particles $\beta=\sqrt{1-m_S^2/m_\chi^2}$ becomes small, and thus the
Mandelstam variables satisfy $t\simeq u$, causing $\mathcal
M_t\mathcal \simeq M_u$. Their coherent addition enhances the
total annihilation amplitude, with the interference term
$2\,\mathrm{Re}(\mathcal M_t\mathcal M_u^\ast)$ providing its largest
positive contribution, enhancing the total amplitude squared. This
leads to a significant reduction of the relic abundance near the
threshold region.

Further, note that the primary role of the RHN is to provide additional depletion channels for dark fermion, making it easier to obtain the observed relic density over a wider region of parameter space. As $m_R$ increases, the channels in Eq.~\eqref{eqN} become phase-space suppressed, and the scalar-sector processes in Eq.~\eqref{eqS} play an increasingly important role. In the absence of RHN, or when the RHN is very heavy, the relic density is mainly determined by the scalar sector interactions shown in Eq.~\eqref{eqS} are relevant. In that case, obtaining the correct relic abundance typically requires $m_\chi\sim m_S$ so that annihilation into scalar pairs is efficient.

%% file: signatures.tex
\section{Multi-messenger probes}
\label{section4:signatures}
The $\mathbb{Z}_3$ symmetry operative in our scenario renders collider
signals very unlikely as the cross section for DM particle production
is highly suppressed. On the other hand, the interplay between the
thermal history of the dark sector and the SFOEWPT gives rise to
several observable signatures through which the viable parameter space
of the model can be probed. The most prominent of these are the
gravitational wave observations, indirect detection searches, and
direct detection prospects. In this section, we discuss these in turn.

\subsection{Gravitational wave signatures}
During the two-step phase transition (a SFOEWPT) discussed in the
previous section, the Universe evolves from the intermediate
$\mathbb{Z}_3$-broken vacuum to the electroweak vacuum through the
nucleation and expansion of true-vacuum bubbles. The associated bubble
collisions, sound waves, and magnetohydrodynamic (MHD) turbulence in
the primordial plasma generate a stochastic gravitational wave (GW)
background.

The GW spectrum generated by a cosmological first-order phase
transition is primarily determined by three macroscopic parameters:
the bubble nucleation temperature $T_n$, the released vacuum (latent)
energy at the nucleation temperature $\alpha$, and the inverse
duration of the phase transition normalized to the Hubble expansion
rate $\beta/H$. Together, these parameters determine the amplitude,
peak frequency, and spectral shape of the resulting GW signal.

The bubble nucleation temperature is defined as that when the
nucleation rate of critical bubbles becomes sufficiently large for the
phase transition to proceed. Quantitatively, it is determined by the
condition~\cite{Linde:1981zj,Mazumdar:2018dfl}
\begin{equation}
\Gamma(T_n)\simeq H^4(T_n),
\qquad
\Longleftrightarrow
\qquad
\frac{S_3(T_n)}{T_n}\simeq 140,
\end{equation}
where $\Gamma(T)$ is the bubble nucleation rate per unit
volume~\cite{Grojean:2006bp,Linde:1981zj}, $H(T)$ is the Hubble
expansion rate, and $S_3(T)$ is the three-dimensional Euclidean bounce
action~\cite{Linde:1981zj}. The second relation provides an excellent
approximation for EWPT. The released vacuum (latent) energy is
quantified by the parameter~\cite{Kamionkowski:1993fg,Kehayias:2009tn}
\begin{equation}
  \alpha_n \equiv \frac{\Delta \rho(T_n)}{\rho_{\rm rad}(T_n)},
  \quad \text{with} \quad \rho_{\rm rad}(T_n)=\frac{\pi^2g_\ast}{30}\,T_n^4,
\end{equation}
with $\rho_{\rm rad}$ denoting the energy density stored in radiation
({\em i.e.} relativistic species) and $g_\ast$ the effective number of
relativistic degrees of freedom at $T_n$. Similarly,
\begin{equation}
\Delta \rho(T_n)\, {\equiv}
\left[
\Delta V_{\rm eff}(T)
-
T\frac{\partial \Delta V_{\rm eff}(T)}{\partial T}
\right]_{T=T_n} \ ,
\end{equation}
 with $\Delta V_{\rm eff}(T)\equiv V_{\rm eff}^{\rm false}(T)-V_{\rm
   eff}^{\rm true}(T)$ denoting the difference in the
 finite-temperature effective potential (see Eq.~\eqref{eq:effpot})
 between the false and true vacuum states. Finally, the inverse
 duration of the phase transition is characterized
 by~\cite{Nicolis:2003tg}
\begin{equation}
\frac{\beta}{H_n}
=
\left.
T\frac{d(S_3/T)}{dT}
\right|_{T=T_n},
\end{equation}
Physically, $\beta^{-1}$ represents the characteristic time scale of
the phase transition. Consequently, a smaller value of $\beta/H$
corresponds to a longer-lasting phase transition, which generally
enhances the amplitude of the resulting stochastic GW signal.

\begin{figure}
    \centering
    \includegraphics[width=0.48\linewidth]{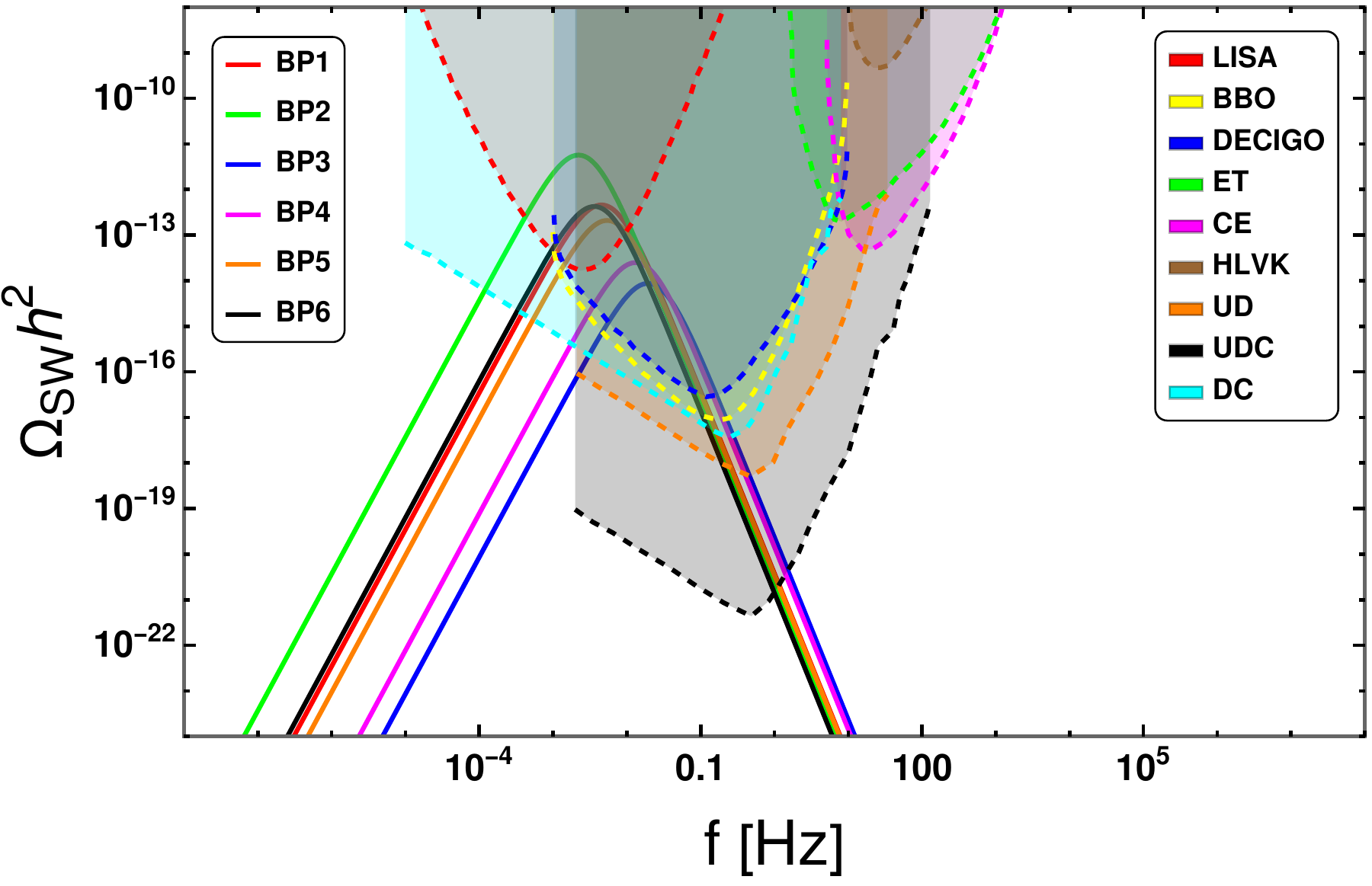}
    \includegraphics[width=0.48\linewidth]{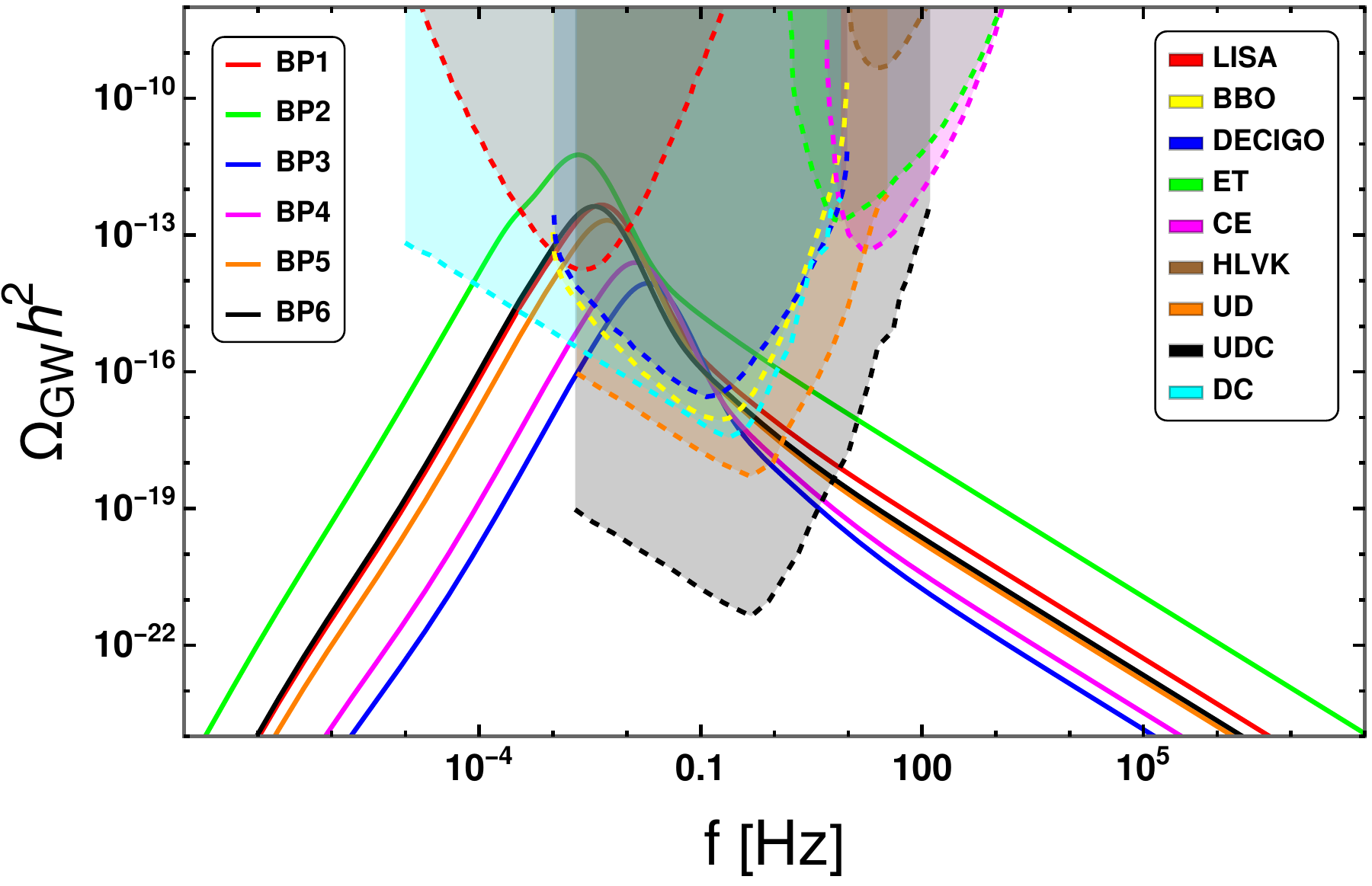}
   \caption{Gravitational wave spectrum produced during the
     phase transition for different benchmark points. The left panel
     shows the contribution from sound waves, while the right panel
     presents the total GW spectrum. Sensitivity
     expectations for future GW detectors, including
     LISA~\cite{LISA:2017pwj}, BBO~\cite{Yagi:2011wg}, the Einstein
     Telescope (ET)~\cite{Punturo:2010zz}, Cosmic Explorer
     (CE)~\cite{Reitze:2019iox}, the HLVK network (LIGO
     Hanford--Livingston, Virgo, and
     KAGRA)~\cite{LIGOScientific:2014pky,VIRGO:2014yos,KAGRA:2018plz},
     DECIGO, Ultimate DECIGO (UD), DECIGO Correlation (DC), and
     Ultimate DECIGO Correlation (UDC)~\cite{Nakayama:2009ce}, are
     given by the shaded regions.}
    \label{fig:gwplot}
\end{figure}

The stochastic GW energy density spectrum receives three main
contributions arising from bubble wall collisions, sound waves, and
magnetohydrodynamic turbulence. Thus, the individual contributions depend additionally on the bubble wall velocity and the corresponding efficiency factors. The total GW energy density spectrum
can therefore be approximated as~\cite{Ellis:2020awk,Caprini:2015zlo}
\begin{equation}\label{eq:GWTotal}
\Omega_{\rm GW}h^2 \approx \Omega_{\rm col} h^2 + \Omega_{\rm sw} h^2 + \Omega_{\rm tur} h^2,
\end{equation}
where $\Omega_{\rm col}$, $\Omega_{\rm sw}$, and $\Omega_{\rm tur}$
denote the contributions from bubble wall collisions, sound waves, and
MHD turbulence, respectively~\cite{Caprini:2015zlo}. Here, $h \equiv
H_0/(100\,\mathrm{km\,s^{-1}\,Mpc^{-1}})$, with $H_0$ being the
present-day Hubble constant~\cite{DES:2017txv}. The explicit
expressions for these individual contributions can be found in
Ref.~\cite{Das:2026zuo}. The dominant contribution arises from sound waves generated in the plasma after bubble percolation~\cite{Hindmarsh:2013xza,Hindmarsh:2016lnk,Hindmarsh:2017gnf}, as shown in Fig.\,\ref{fig:gwplot}, with the corresponding GW energy-density spectrum given by~\cite{Das:2026zuo}
\begin{equation}\label{eq:GWswpart}
\Omega_{\rm sw} h^2 = 2.65 \times 10^{-6}\; \Upsilon(\tau_{\rm sw}) \left(  \frac{\beta}{H_n} \right)^{-1} v_w \left( \frac{\kappa_{\rm sw} \alpha_n}{1 + \alpha_n} \right)^2 \left( \frac{100}{g^{\ast}} \right)^{1/3} \left( \frac{f}{f_{\rm sw}} \right)^3 \left[ \frac{7}{4 + 3 \left( f/f_{\rm sw} \right)^2} \right]^{7/2},
\end{equation}
where $\kappa_{\rm sw}$ denotes the efficiency factor for the conversion of latent heat into bulk fluid motion~\cite{Kamionkowski:1993fg}, while the peak frequency
\begin{equation}\label{eq:PF2}
f_{\rm sw} = 1.9 \times 10^{-5} \left( \frac{1}{v_w} \right) \left( \frac{\beta}{H_n} \right) \left( \frac{T_n}{100 \, {\rm GeV}} \right) \left( \frac{g^{\ast}}{100} \right)^{1/6} \, {\rm Hz}.
\end{equation}
The factor $\Upsilon(\tau_{\rm sw})$ accounts for the finite sound-wave lifetime~\cite{Hindmarsh:2017gnf}, while we take the relativistic wall velocity $v_w=1$~\cite{Kamionkowski:1993fg,Espinosa:2010hh}.

For the benchmark points listed in Tab.\,\ref{tab:FOPT}, we compute the
phase transition parameters relevant for EWPT dynamics and the
resulting stochastic GW spectra using the publicly available code
\texttt{CosmoTransitions}~\cite{Wainwright:2011kj}. Fig.\,\ref{fig:gwplot} shows the stochastic GW spectra corresponding
to the six benchmark points together with the projected sensitivities
of future space-based GW observatories mentioned in the caption. Among
all the BPs, BP2 yields the largest GW amplitude owing to its relatively
large $\alpha_n$, while maintaining a moderate value of
$\beta/H_n$ as expected from Eq.\,\eqref{eq:GWswpart}. Consequently, its peak lies well within the projected
sensitivity of LISA. Similarly, the GW signals corresponding to BP1,
BP5, and BP6 also fall within the LISA sensitivity band, although with
comparatively smaller amplitudes due to their relatively weaker phase
transitions. In contrast, BP3 and BP4 are characterized by smaller
values of $\alpha$ together with larger values of $\beta/H_n$,
resulting in weaker GW signals with peak frequencies shifted to higher
values. Consequently, these benchmark points are more effectively
probed by DECIGO rather than LISA.

\subsection{Indirect detection signatures}

The parameter region with $m_N<m_\chi$ can be probed through gamma-ray
observations from dark matter annihilation. In the present model, both
the annihilation process and the semi-annihilation process contribute to the gamma-ray
signal. The unstable RHNs subsequently decay into Standard Model particles, whose hadronization and radiative decays produce a continuum of gamma rays. However, we find that the limits derived from the annihilation channel are considerably
stronger over the parameter space of interest. This is primarily because the
annihilation process produces two RHNs in the final state, both of
which subsequently decay into Standard Model particles, resulting in a
significantly larger gamma-ray yield per annihilation than in the
semi-annihilation channel, where only a single RHN is produced. Therefore, in the
following, we focus on the constraints arising from
$\bar{\chi}\chi\rightarrow NN$.

\begin{figure}[!h]
    \centering
    \includegraphics[width=0.49\linewidth]{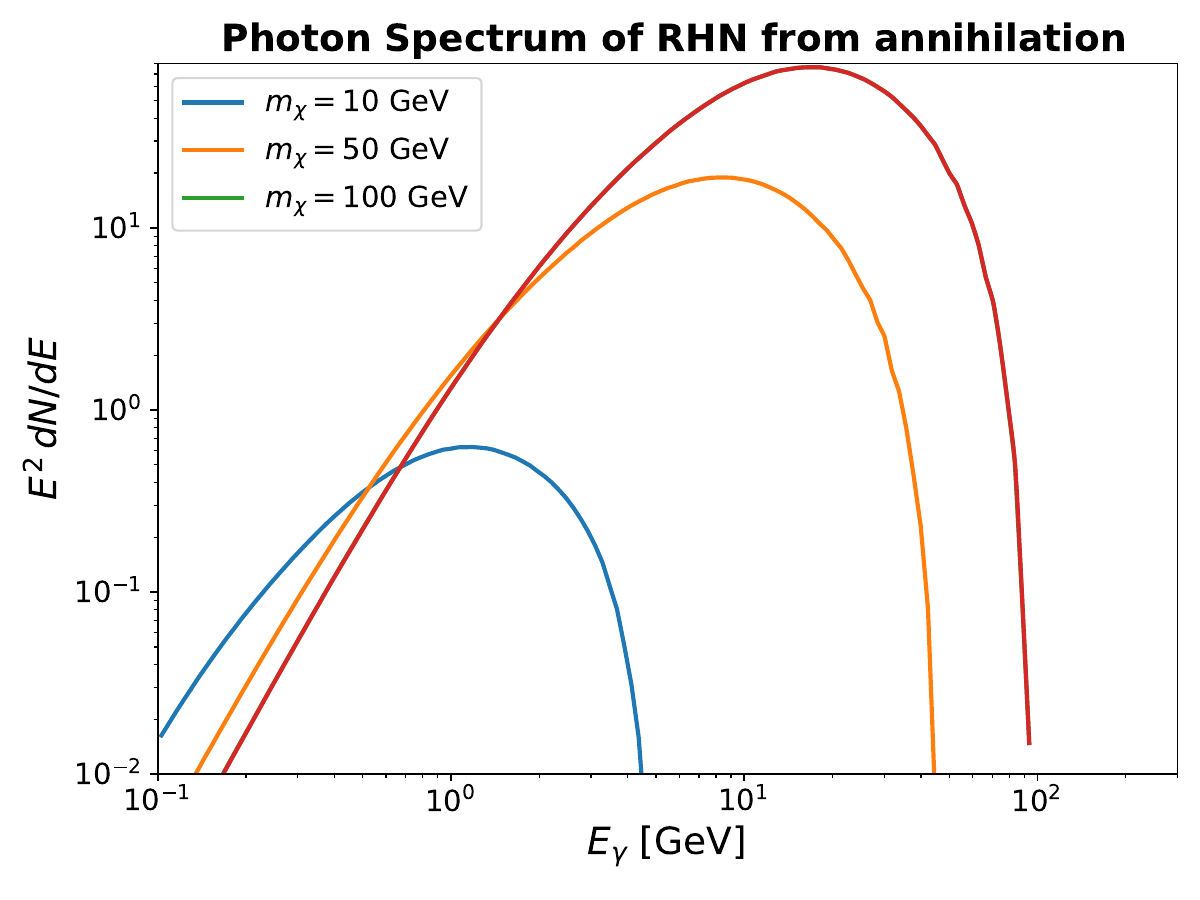}
    \includegraphics[width=0.49\linewidth]{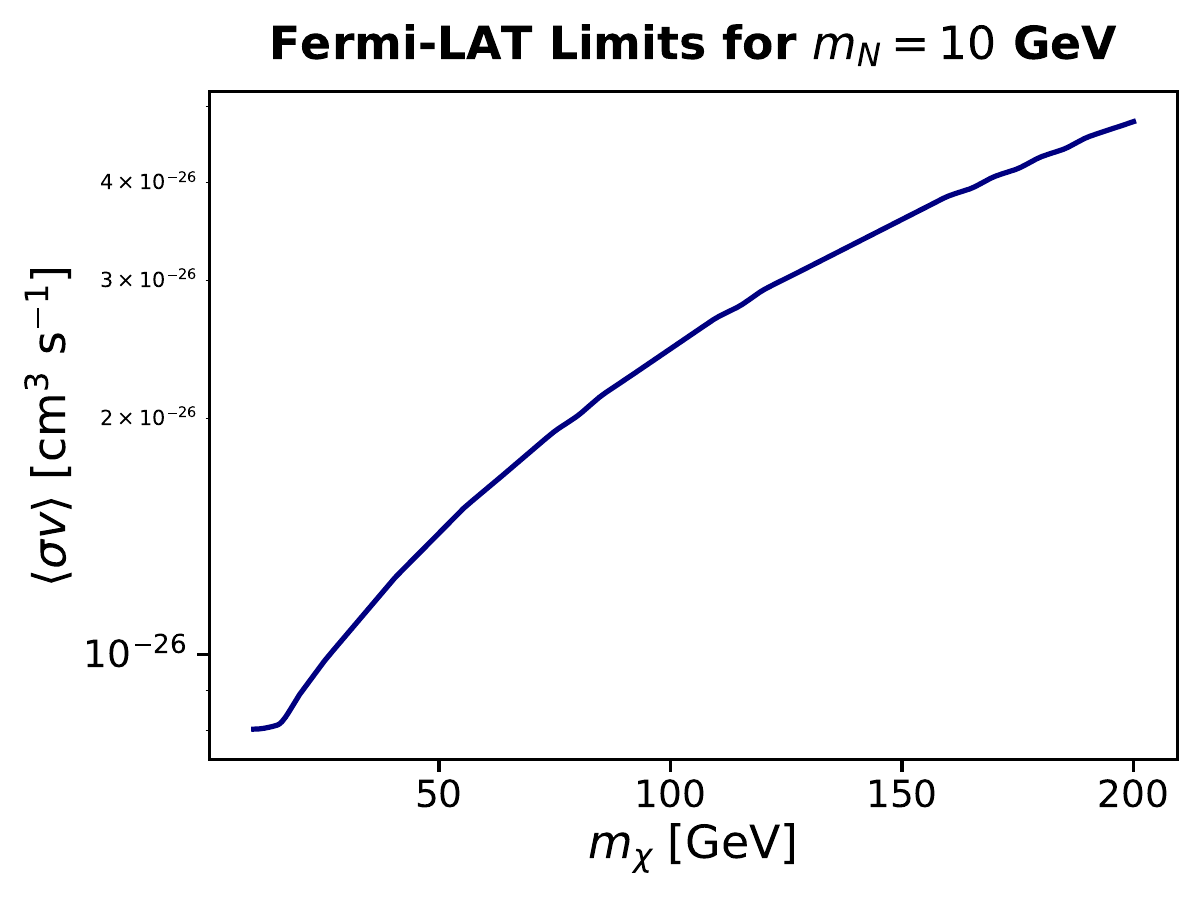}
    \caption{Left panel: Gamma-ray spectra from the annihilation
      process $\chi\chi \rightarrow N N$ for representative dark
      matter masses. Right panel: Corresponding $95\%$ C.L. upper
      limits on $\langle \sigma v \rangle_{\chi\chi\to N N}$ derived
      from the {\it Fermi}-LAT dwarf spheroidal galaxy likelihood
      data.}
    \label{fig:indirectDet}
\end{figure}

For a benchmark RHN mass of $m_R=10$ GeV, the prompt gamma-ray
spectrum from $\bar{\chi}\chi\rightarrow NN$ is generated using {\tt
  Pythia~8.3.1}~\cite{Bierlich:2022pfr}. The resulting spectra, shown in
Fig.~\ref{fig:indirectDet} in terms of $E_\gamma^2
dN_\gamma/dE_\gamma$, exhibit the expected hardening with increasing
dark matter mass due to the larger boost of the produced RHNs.

To derive the corresponding indirect detection limits, we employ the
publicly available bin-by-bin likelihood data released by the {\it
  Fermi}-LAT Collaboration for dwarf spheroidal
galaxies~\cite{PhysRevLett.115.231301}. For a given annihilation cross
section $\langle\sigma v\rangle$, the differential gamma-ray flux from
a dwarf galaxy is given by
\begin{equation}
\frac{d\Phi_\gamma}{dE_\gamma}
=
\frac{\langle\sigma v\rangle}{8\pi m_\chi^2}
\frac{dN_\gamma}{dE_\gamma}
J,
\end{equation}
where $dN_\gamma/dE_\gamma$ is the photon spectrum per annihilation obtained from {\tt Pythia}, and
\begin{equation}
J=\int_{\Delta\Omega} d\Omega
\int_{\rm l.o.s.}\rho^2(l,\Omega)\,dl
\end{equation}
is the astrophysical $J$-factor, obtained by integrating the squared dark matter density along the line of sight over the observed solid angle.

The predicted flux in each energy bin is compared with the tabulated
{\it Fermi}-LAT likelihood profiles by interpolation, and the total
likelihood is constructed by summing the log-likelihood contributions
over all energy bins and all dwarf spheroidal galaxies included in the {\it Fermi}-LAT analysis. For each dark matter mass, we scan
$\langle\sigma v\rangle$ and determine the corresponding test
statistic
\begin{equation}
\Delta(-2\ln\mathcal L)
=
-2\ln\mathcal L
+
2\ln\mathcal L_{\rm min},
\end{equation}
where $\mathcal L_{\rm min}$ denotes the maximum likelihood. Since the
annihilation cross section is the only parameter of interest for a
fixed $m_\chi$, the test statistic is assumed to follow a $\chi^2$
distribution with one degree of freedom according to Wilks'
theorem. The 95\% confidence level upper limit on the annihilation
cross section is obtained by imposing $\Delta(-2\ln\mathcal L)=2.71.$
The resulting upper limits on $\langle\sigma v\rangle$ are shown in
Fig.~\ref{fig:indirectDet}.

\begin{figure}[!h]
    \centering
    \includegraphics[width=0.49\linewidth]{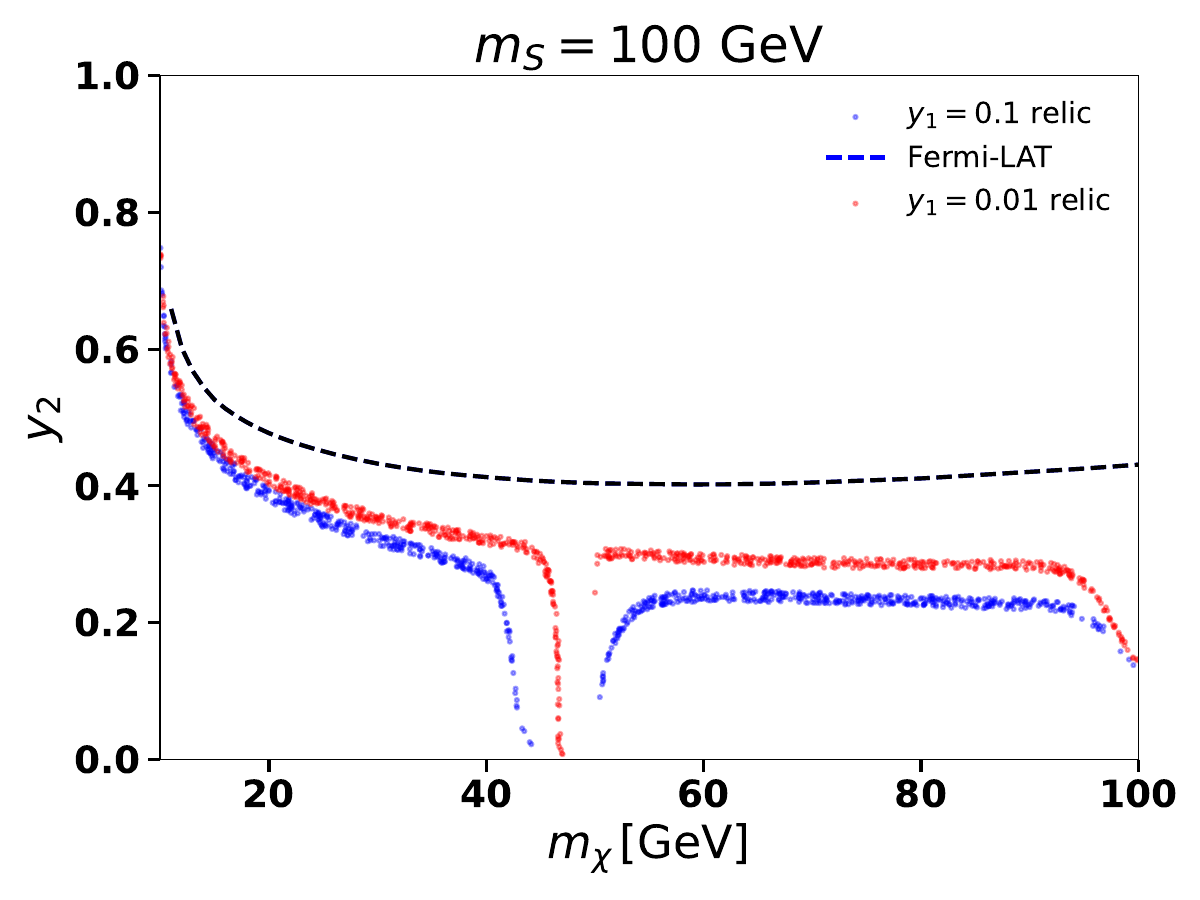}
    \includegraphics[width=0.49\linewidth]{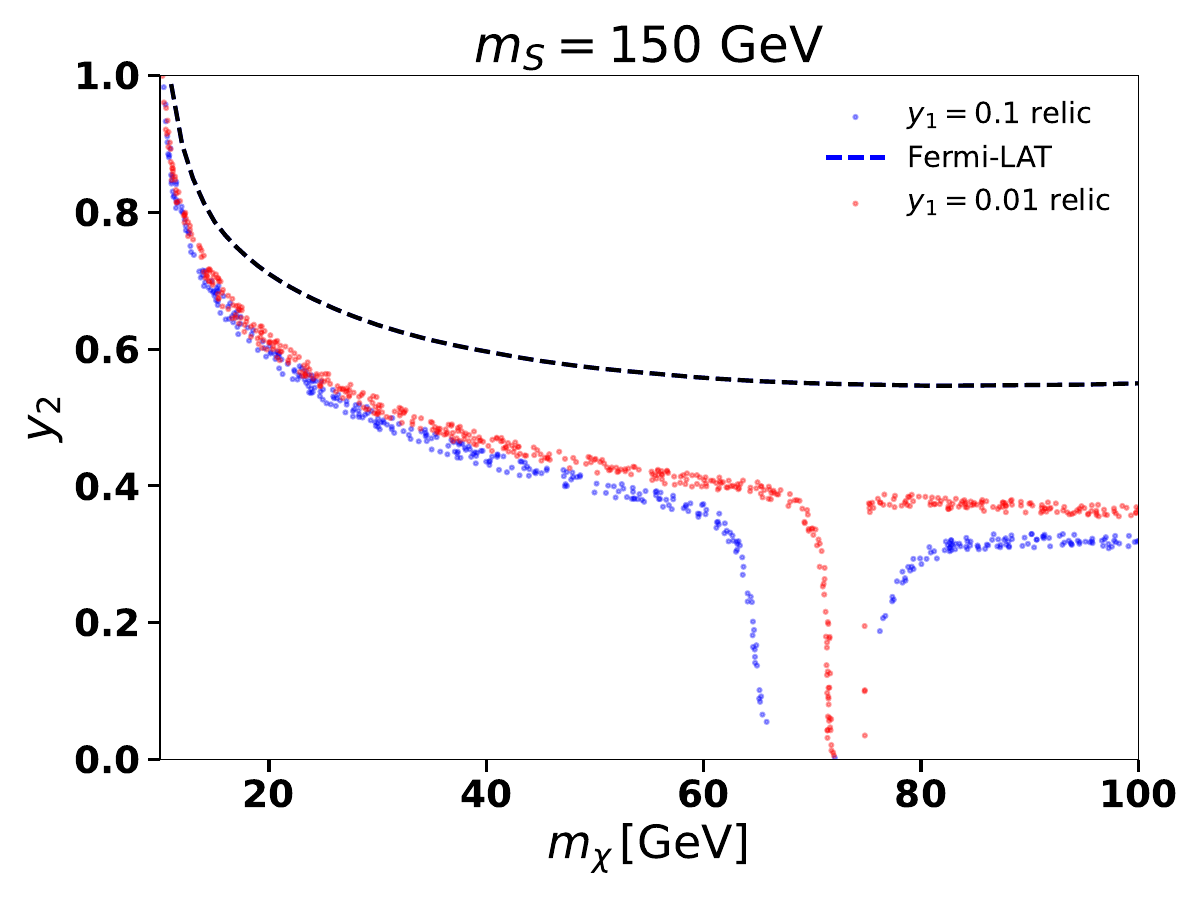}
    \caption{Comparison of the relic-density favored regions with the
      95\% C.L. Fermi-LAT limits in the $m_\chi-y_2$ plane for
      $m_S=100~(150)$ GeV. The blue and red dotted points satisfy the
      observed dark matter relic abundance, while the region above
      dashed curves is
        ruled out on account of the $\bar{\chi}\chi\to NN$
      annihilation channel failing the Fermi-LAT constraints.}
    \label{fig:compare}
\end{figure}

The impact of these limits on the model parameter space is illustrated
in Fig.~\ref{fig:compare}, where the Fermi-LAT exclusion contours are
overlaid on the regions satisfying the observed dark matter relic
abundance. This comparison demonstrates the extent to which the
indirect detection constraints probe the viable parameter space of the
model.

 \begin{figure}
     \centering
     \includegraphics[width=0.49\linewidth]{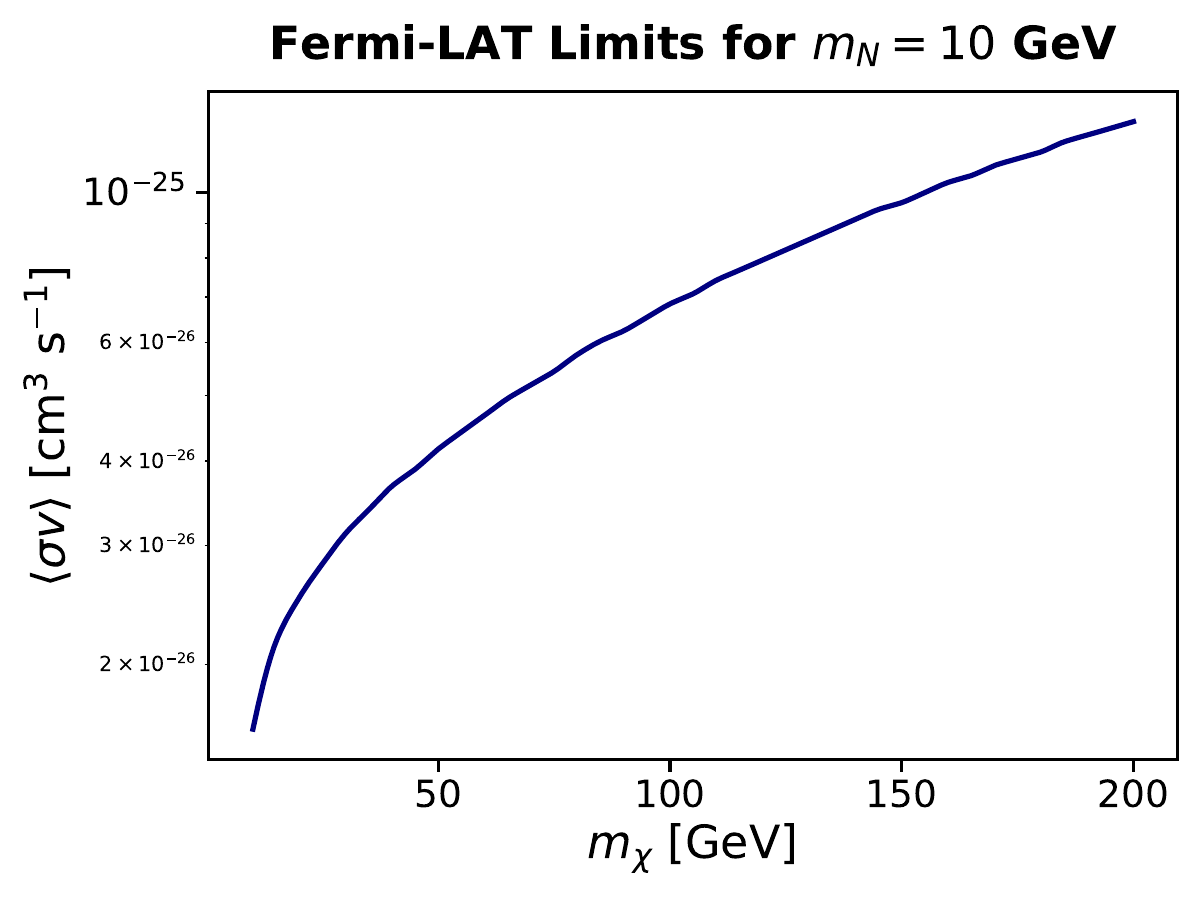}
     \caption{ $95\%$ C.L. upper limits on the $\langle \sigma v \rangle_{\bar{\chi}\chi\to\chi N}$ derived from the {\it Fermi}-LAT dwarf spheroidal galaxy likelihood data.}
     \label{fig:XXXN}
 \end{figure}

An interesting exception to the above discussion occurs in the
vicinity of $m_\chi\simeq m_S/2$, where the semi-annihilation channel
is resonantly enhanced. And since this process proceeds through an
$s$-channel scalar and is dominated by the $s$-wave contribution, its
annihilation cross section today is expected to be comparable to that
at freeze-out. However, even in this resonant region, the current
Fermi-LAT upper limits remain above the thermal relic cross section,
and hence no additional parameter space is excluded (see
Fig.\ref{fig:XXXN}). While the present analysis focuses on a light RHN
benchmark, the corresponding indirect detection prospects for heavier
RHNs have been investigated in Ref.~\cite{Bandyopadhyay:2022tsf}.

\subsection{Direct detection signatures}
\begin{figure}[!htb]
\centering
\begin{tikzpicture}[scale=0.9]

\begin{scope}
\coordinate (A) at (-0.60,1.30);
\coordinate (B) at (0.60,1.30);
\coordinate (C) at (0.00,0.25);
\draw[fermion] (A)--(B);
\node at (0.00,1.63){$\chi$};
\draw[dashed] (B)--(C);
\node[right] at (0.45,0.75){$S$};
\draw[dashed] (C)--(A);
\node[left] at (-0.45,0.75){$S$};
\draw[fermion] (-0.85,1.55)--(A);
\draw[antifermion] (0.85,1.55)--(B);
\node at (-1.00,1.72){$\chi$};
\node at (1.00,1.72){$\chi$};
\draw[dashed] (C)--(0,-0.70);
\node[left] at (-0.12,-0.15){$h$};
\draw[antifermion] (0,-0.70)--(-0.80,-1.45);
\draw[fermion] (0,-0.70)--(0.80,-1.45);
\node at (-0.95,-1.62){$q$};
\node at (0.95,-1.62){$q$};
\end{scope}
\begin{scope}[xshift=5.2cm]
\coordinate (A) at (-0.60,1.30);
\coordinate (B) at (0.60,1.30);
\coordinate (C) at (0.00,0.25);
\draw[fermion] (A)--(B);
\node at (0.00,1.63){$N$};
\draw[dashed] (B)--(C);
\node[right] at (0.45,0.75){$S$};
\draw[dashed] (C)--(A);
\node[left] at (-0.45,0.75){$S$};
\draw[fermion] (-0.85,1.55)--(A);
\draw[antifermion] (0.85,1.55)--(B);
\node at (-1.00,1.72){$\chi$};
\node at (1.00,1.72){$\chi$};
\draw[dashed] (C)--(0,-0.70);
\node[left] at (-0.12,-0.15){$h$};
\draw[antifermion] (0,-0.70)--(-0.80,-1.45);
\draw[fermion] (0,-0.70)--(0.80,-1.45);
\node at (-0.95,-1.62){$q$};
\node at (0.95,-1.62){$q$};
\end{scope}
\end{tikzpicture}
\caption{One-loop Feynman diagrams contributing to the spin-independent DM--nucleon direct detection cross section.}
\label{fig:one-loopDD}
\end{figure}
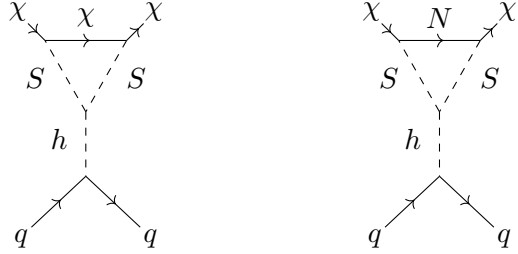

The fermionic DM candidate does not induce a tree-level
spin-independent scattering cross section. However, an observable
spin-independent signal arises at one loop through the exchange of the
scalar as well as RHN mediator, as shown in
Fig.\,\ref{fig:one-loopDD}.

The one-loop vertex correction amplitude corresponding to the diagrams
in Fig.\,\ref{fig:one-loopDD}, which contributes to the
spin-independent DM--nucleon direct detection cross section, is given
by~\cite{Maleki:2022zuw,Okada:2013rha}
\begin{align}
\mathcal{M}
=&\;
\frac{m_q c_{hss}}
{16\pi^{2} v_h m_h^{2}}
\Bigg[(2y_{1})^{2}\,G_\chi(m_S,m_\chi) + y_{2}^{2}\,G_N(m_S,m_\chi,m_R)
\Bigg]
\,\bar{q}q\,\bar{\chi}\chi, \quad c_{hss} =  v_h \lambda_{hs},
\end{align}
where the loop functions $G_{\chi,N}$, of mass dimension $-1$, are
given in App.\,\ref{app:couplings}. Therefore, the spin-independent
cross section at one-loop level is given by
\begin{eqnarray}
    \sigma_{\rm{SI}} =\frac{m_p^4 f_N^2 m_\chi^2 }{\pi(m_p+m_\chi)^2} \left(\frac{ c_{hss} }
{16\pi^{2} v_h m_h^{2}}\right)^2\Big|(2y_1)^2G_\chi(m_S,m_\chi)
+ y_2^2G_N(m_S,m_\chi,m_R)
 \Big|^2
\end{eqnarray}
where $m_p$ denotes the nucleon mass and $f_N$ is the effective
Higgs--nucleon coupling. For numerical analysis, we adopt
$m_p=0.946$~GeV~\cite{Alanne:2018zjm} and
$f_N\simeq0.3$~\cite{Alarcon:2011zs,Alarcon:2012nr,Cline:2013gha}.

\begin{figure}
    \centering
    \includegraphics[width=0.49\linewidth]{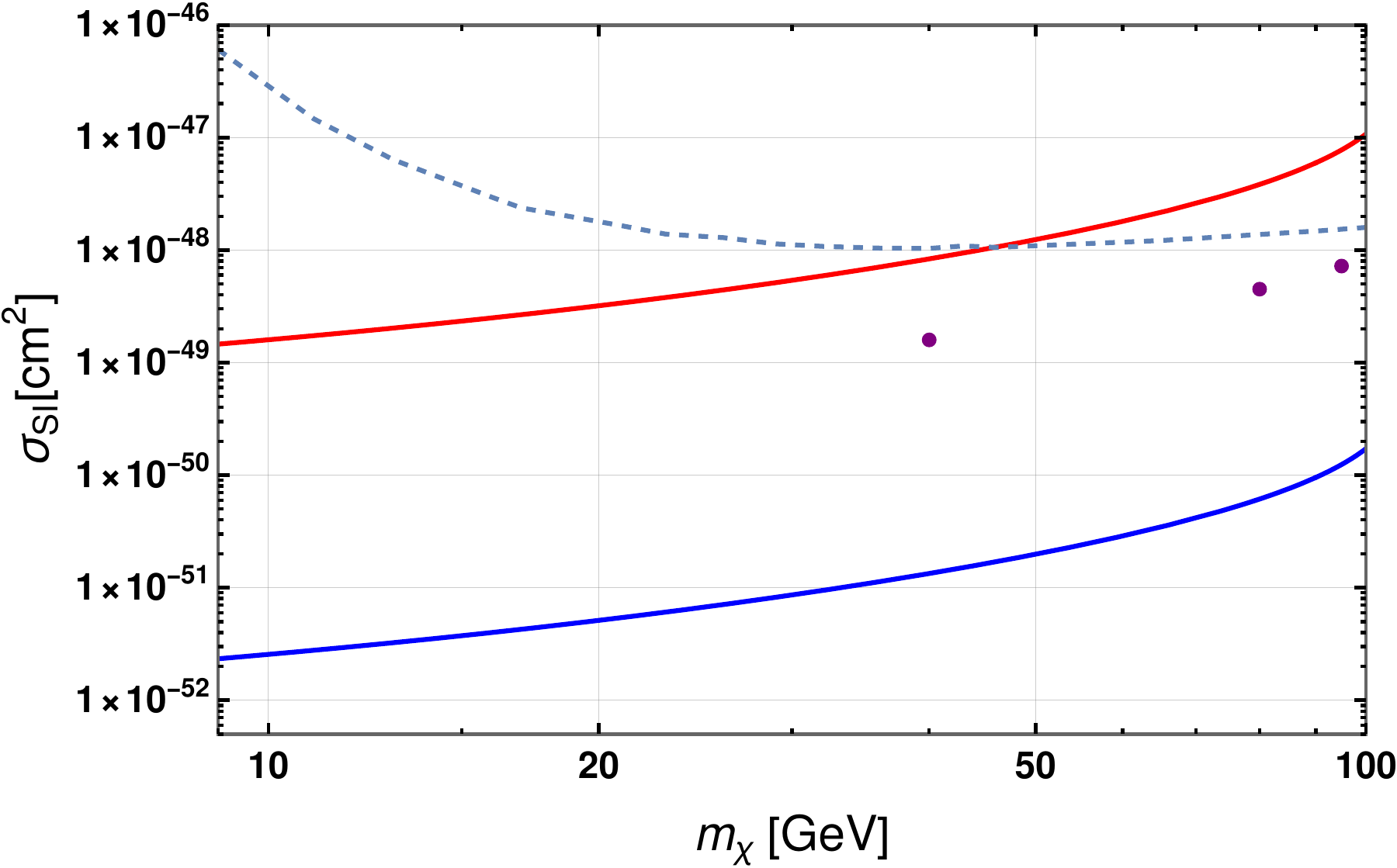}
    \includegraphics[width=0.49\linewidth]{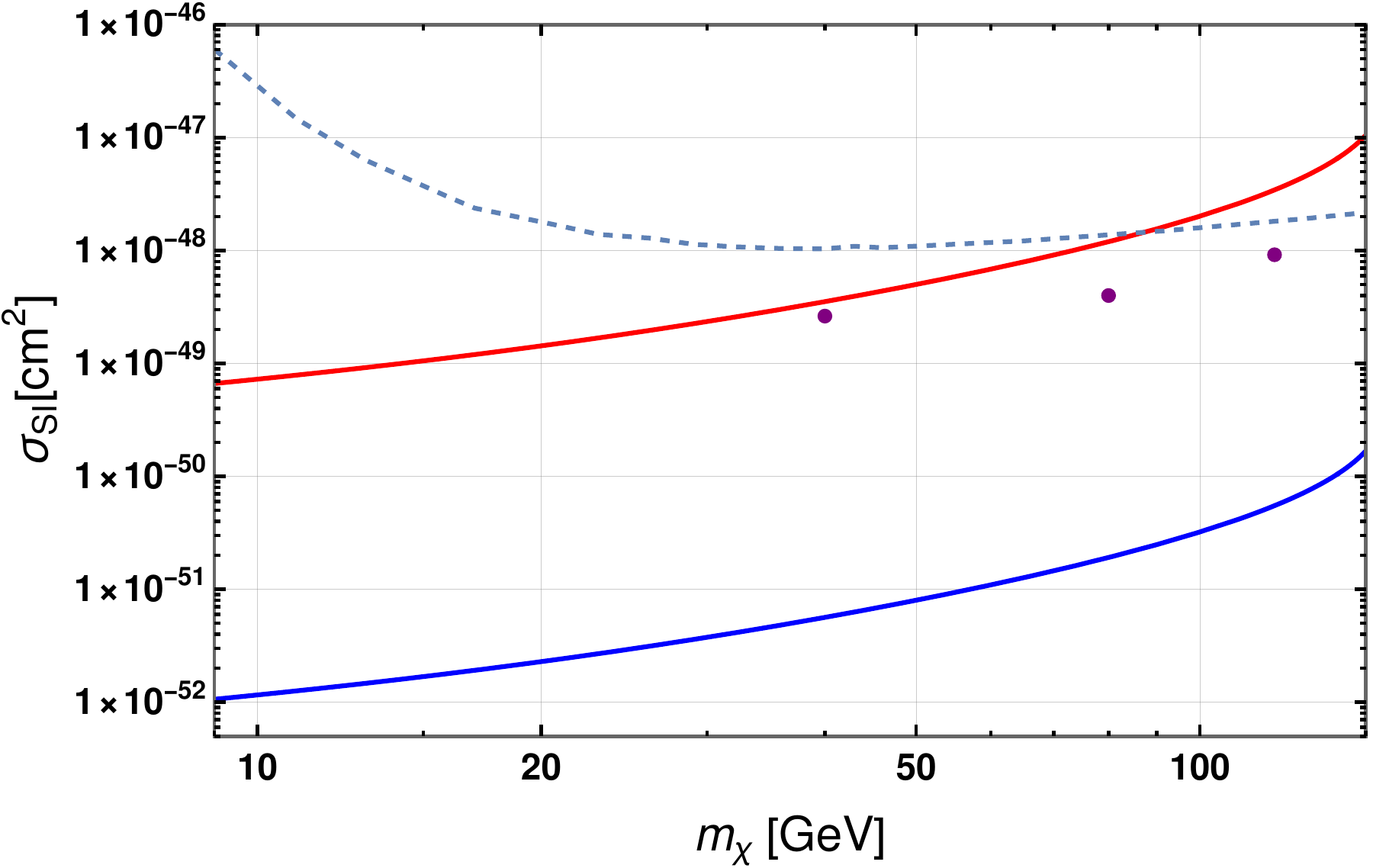}
    \caption{Variation of the one-loop spin-independent DM--nucleon scattering cross section with the DM mass. The red (blue) curves correspond to $y_2=0.5\,(0.1)$ with $m_R=10~\mathrm{GeV}$. The left (right) panel shows $\lambda_{hs}=1.0$ and $m_S=100~\mathrm{GeV}$ ($\lambda_{hs}=1.5$ and $m_S=150~\mathrm{GeV}$). The dashed curve denotes the latest spin-independent direct-detection limit from LZ~\cite{LZ:2024zvo}. The purple dots indicate the six benchmark points.}
    \label{fig:DD}
\end{figure}

The resulting spin-independent cross section is shown in
Fig.\,\ref{fig:DD} for the model parameters. These are obtained for
$y_1<y_2$. Since the relic abundance depends only weakly on $y_1$,
this choice allows us to isolate the RHN-mediated contribution
proportional to $y_2$ and illustrate its impact more
transparently. Increasing $y_1$ to values comparable to $y_2$ enhances
the contribution from the $\chi$-mediated diagram and introduces
interference between the two amplitudes, resulting in a moderately
enhanced spin-independent cross section. However, the behaviour of the
constraints and the overall pattern of the allowed parameter space
remain unchanged.

For a relatively large Yukawa coupling, $y_2=0.5$, a significant fraction of the parameter space corresponding to larger DM masses is already excluded by the latest LZ limits. However, for fixed Yukawa couplings, the constraints become weaker as the scalar and RHN masses increase owing to the suppression of the loop functions. Consequently, the excluded parameter space becomes smaller for heavier scalar masses, despite the slightly larger Higgs portal coupling, indicating that the dependence on the portal coupling is relatively mild. Reducing $y_2$ further suppresses the loop contribution, thereby weakening the direct detection bounds. Thus, the benchmark scenario with $m_N\simeq m_S$, which typically corresponds to a moderately smaller value of $y_2$, remains viable even for relatively heavy dark matter masses and
constitutes a target for future direct detection
experiments.

We finally comment on the possibility of observing the scalar sector directly. Since the scalar is responsible for realizing the SFOEWPT, the required Higgs portal coupling is relatively large, leading to an enhanced spin-independent scattering cross section. Consequently, a stable scalar dark matter component is excluded over most of the parameter space. A notable exception occurs in the vicinity of the Higgs
resonance, $m_S\simeq m_h/2$, where resonant annihilation through the
Higgs portal efficiently depletes the scalar relic abundance. If, in addition, the spectrum is sufficiently compressed,
$m_\chi\simeq m_S$, the scalar can survive as a long-lived
sub-dominant component while the fermion continues to account for the dominant relic density. Although this region is highly restricted, it remains compatible with current direct detection limits and provides an additional channel for future direct detection experiments.

%% file: conclusion.tex
\section{Summary and Conclusion}
\label{section5:conclusion}
The astrophysical and cosmological evidence for DM motivates the
exploration of physics beyond the SM to accommodate viable DM
candidates. In this work, we have extended the SM by introducing a
Dirac fermion, $\chi$, a right-handed neutrino, $N_R$, and a complex
scalar singlet, $S$. The Dirac fermion $\chi$ and the scalar $S$ carry
non-trivial and equal $\mathbb{Z}_3$ charges.
In contrast, the RHN together with all SM fields, is neutral under the
$\mathbb{Z}_3$ symmetry and transforms trivially. In this framework,
we analyse the thermal history, which contains distinct phases
characterized by the breaking patterns of the electroweak and
$\mathbb{Z}_3$ symmetries. At very high temperatures, presumably both
are unbroken, while at low temperatures the electroweak symmetry is
broken and the $\mathbb{Z}_3$ symmetry is preserved. An intermediate
phase may exist in which the electroweak symmetry remains unbroken
while the $\mathbb{Z}_3$ symmetry is spontaneously broken.

During the $\mathbb{Z}_3$-broken phase, the scalar field $S$ acquires
a non-zero VEV due to the large Higgs portal coupling, while the Higgs
field retains a vanishing VEV. The non-zero VEV of $S$
induces mixing between the Dirac fermion $\chi$ and the RHN $N_R$,
splitting the Dirac fermion into two pseudo-Dirac states, one CP-even
and the other CP-odd.  For the parameter space that simultaneously
satisfies the observed DM relic abundance and realizes a SFOEWPT, all
dark-sector particles remain relativistic or only mildly
non-relativistic throughout the $\mathbb{Z}_3$-broken phase. The DM
freeze-out occurs only after the completion of the EWPT, when $\langle S\rangle$ vanishes, the $\mathbb{Z}_3$
symmetry is restored, and the standard thermal freeze-out mechanism
determines the present-day relic abundance.

We demonstrate that the observed DM relic abundance and a SFOEWPT can
be simultaneously realized for a wide swathe of the
  parameter space. Our analysis further shows that achieving a
SFOEWPT requires a larger Higgs portal coupling as the scalar mass
increases, indicating a positive correlation between the scalar mass
and the strength of the portal interaction necessary to satisfy both
cosmological and DM constraints.

The RHN $N_R$ plays a crucial role in the DM phenomenology by
significantly influencing the DM relic abundance through its
interactions with the dark sector.  The Yukawa interaction
$S\,\bar{\chi}\,N_R$, together with the cubic scalar interaction
permitted by the $\mathbb{Z}_3$ symmetry, gives rise to characteristic
semi-annihilation channels that are absent in conventional
$\mathbb{Z}_2$ symmetric DM models, enabling the observed DM relic
abundance over a broad region of the parameter space shown in
Figs.\,\ref{fig:relicN_band} and \ref{fig:relicN_band100}.

Of particular interest are
the stochastic GW signals generated during the
SFOEWPT. The GW spectrum is predominantly
sourced by long-lasting sound waves in the plasma, while the
contribution from magnetohydrodynamic turbulence is subdominant and
the bubble-collision contribution is negligible under the assumption
of non-runaway bubble walls. The predicted GW spectra shown in
Fig.\,\ref{fig:gwplot} peak in the mHz--sub-Hz frequency range,
placing a significant portion of the signal within the projected
sensitivities of future space-based GW observatories such as LISA,
BBO, and DECIGO. These results highlight the complementarity between
GW observations and DM searches in probing the parameter space of the
model.

Since $S$ does not acquire a VEV after the EWPT, it
does not mix with the SM Higgs boson. Therefore, the fermionic DM
candidate has no tree-level interaction with nucleons. However,
scalar-mediated interactions generate a loop-level spin-independent
direct detection signal. We also investigated the indirect detection
prospects of the model using the latest {\it Fermi}-LAT dwarf
spheroidal galaxy observations. Current Fermi-LAT limits are
consistent with the viable parameter space, while future gamma-ray
observations will further probe the model.

Our analysis shows that the scalar $S$ cannot constitute a viable dominant DM
component over most of the parameter space because the relatively
large Higgs portal coupling required to realize the SFOEWPT
simultaneously enhances the spin-independent DD cross section to
  phenomenologically unacceptable levels. The only viable region
corresponds to the Higgs resonance, $m_S \simeq m_h/2$, where resonant
annihilation through the Higgs portal suppresses the scalar relic
abundance sufficiently to evade present DD constraints. In this
region, $\chi$ remains responsible for the observed DM relic density,
while $S$ contributes only a negligible fraction, making the
Higgs-resonance region an interesting target for future DD
experiments. 

In summary, we have presented a minimal $\mathbb{Z}_3$-symmetric
fermionic DM model that simultaneously addresses the origin of DM and
the dynamics of the EWPT. A distinctive feature of the model is its
non-trivial thermal history, which differs from that of traditional
WIMP scenarios and naturally emerges from the interplay of the
dark-sector interactions. The observed DM relic abundance is
reproduced through the standard thermal freeze-out mechanism via the
interplay of annihilation, semi-annihilation and dark-sector
conversion processes inherent to the $\mathbb{Z}_3$ symmetry. The
model realizes a SFOEWPT over the viable parameter space and predicts
stochastic GW signals with peak frequencies lying within the projected
sensitivities of future space-based detectors such as LISA, BBO, and
DECIGO. The interplay between DM phenomenology, semi-annihilation, the
EWPT, and GW signatures makes this framework a compelling and
experimentally testable extension of the SM.

\section*{Acknowledgement}
DC acknowledges the ANRF, Government of India, for support through the
project CRG/\allowbreak2023/008234 and the IoE, University of Delhi
grant IoE/2025-26/12/FRP. JD acknowledges the ANRF (formerly Science and
Engineering Research Board (SERB)), Government of
India, for the national postdoctoral fellowship (NPDF) grant PDF/2023/0015. DS acknowledges the ANRF for support through the project ANRF/ECRG/2025/002846.

%% file: appendix.tex
\appendix
\section{Thermal Self Energies}\label{app:thermal}
The bosonic and fermionic thermal functions are~\cite{Quiros:1999jp}
\begin{eqnarray}\label{eq:thermalF}
  J_{B/F}\Big(\frac{m^2(h,s)}{T^2}\Big)=\int_0^\infty  dx \, x^2 \log \left(1 \mp e^{-\sqrt{x^2+\frac{m^2(h,s)}{T^2}}}\right) .  
\end{eqnarray}
For $m^2/T^2 \gg 1$, the thermal functions are exponentially suppressed, implying that heavy particles in loop contribute negligibly to the finite-temperature effective potential. In the opposite regime, $x \equiv \frac{m^2}{T^2} \ll 1,$ the thermal functions admit the well-known high-temperature expansions~\cite{Quiros:1999jp}
\begin{eqnarray} \label{eq:expn}
  J_B(x) &= & -\frac{\pi^4}{45} + \frac{\pi^2 x}{12}
  - \frac{\pi}{6} x^{3/2} - \frac{x^2}{32} \ln \frac{x}{a_b } + \cdots,
  \\ 
  J_F(x) &= &  \frac{7\pi^4 }{360} - \frac{\pi^2 x}{24}
     - \frac{x^2}{32} \ln \frac{x}{a_f} + \cdots, 
\end{eqnarray}
where
$a_f = \pi^2\exp \left(3/2-2\gamma_E\right)$ and $a_b = 16 \pi^2\exp \left(3/2-2\gamma_E\right)$,
with $\gamma_E \simeq 0.577$ denoting the Euler--Mascheroni constant.

From the high temperature expansion of the thermal functions, the self-energy correction to the thermal masses of the CP-even scalar can be written as
\begin{align}\label{eq:daisy}
&\Pi_{h h} = \frac{3g^2}{16}  + \frac{g^{\prime 2}}{16}  + \frac{y_q^2}{4} + \frac{y_\ell^2}{12} + \frac{\lambda_h}{2} + \frac{\lambda_{hs}}{12}
,  \\
&\Pi_{ss} = \frac{\lambda_{s}}{3} + \frac{\lambda_{hs}}{6}+ \frac{y_1^2}{3} + \frac{y_2^2}{6} .\label{eq:daisy1}
\end{align}
The thermal resummed masses of the Higgs Goldstone bosons $\chi_i$ are
\begin{eqnarray}
   && \Pi_{\chi_i} = \frac{3g^2}{16}  + \frac{g^{\prime 2}}{16}  + \frac{y_q^2}{4} + \frac{y_\ell^2}{12} + \frac{\lambda_h}{2} + \frac{\lambda_{hs}}{12},\\
   &&\Pi_{\xi} = \frac{\lambda_{s}}{3} + \frac{\lambda_{hs}}{6}+  \frac{y_1^2}{3}+\frac{y_2^2}{16}.
\end{eqnarray}

The longitudinal polarization states of the $W$ bosons receive thermal mass corrections~\cite{Oikonomou:2024jms},
\begin{align}
   & m_{W_L}^2(h) \rightarrow m_{W_L}^2(h,T)= m_{W_L}^2 (h)+ \Pi_{W_L}(T), \quad \Pi_{W_L}(T) = \frac{11}{6}g^2 T^2.
\end{align}
At finite temperature, the longitudinal modes of the $Z$ boson and photon mix, yielding the mass matrix~\cite{Oikonomou:2024jms}
\begin{equation}\label{eq:matrixZ}
\mathcal{M}_{Z_L/\gamma_L}^2(h,T) =
    \begin{pmatrix}
    \frac{1}{4}g^2 h^2 + \frac{11}{6}g^2 T^2 & -\frac{1}{4}g g' h^2\\
    -\frac{1}{4}g g' h^2 & \frac{1}{4}g'^2 h^2 + \frac{11}{6}g'^2 T^2\\
    \end{pmatrix} \, .
\end{equation}
The eigenvalues of this matrix determine the field-dependent masses of the longitudinal polarization states,
\begin{equation}
m^2_{Z_L/\gamma_L}(h,T) = 
(g^2 + g^{\prime 2}) \left(\frac{h^2}{8} + \frac{11}{12}  T^2\right)
  \pm \sqrt{\left(g^2 - g^{\prime 2} \right)^2 \left( \frac{h^2}{8}  + \frac{11T^2}{12}  \right)^2  + \frac{g^2 g^{\prime 2}}{16} h^4 }.
\end{equation}
Expectedly, at $T\neq 0$, the $\gamma_L$ is massive and contributes to the effective potential.

\section{Counter-term potential}\label{app:ct}
The counterterm coefficients are determined by imposing the following renormalization conditions~\cite{Anisha:2022hgv}:
\begin{eqnarray}
 \left\{\frac{d}{dh},\frac{d^2}{dh^2},\frac{d^2}{ds^2}\right\} \Big(V^{\rm CW}_{1-\rm loop}(h,s)+V_{\rm ct}(h,s)\Big) \Big|_{h=v_h,s=0}=0,
 \end{eqnarray}
 where the counterterm potential is defined in Eq.\,\eqref{eq:ct}. 
 Applying the above conditions, the counterterm coefficients are obtained as
\begin{eqnarray}
   && \delta\mu_h^2 =-\frac{1}{2v_h}\left(3 \frac{dV^{\rm CW}}{d h} -v_h \frac{d^2V^{\rm CW}}{d h^2} \right)\Big|_{h=v_h,s=0},\\
  &&  \delta\mu_{s}^2=- \frac{d^2V^{\rm CW}}{d s^2} \Big|_{h=v,s=0}, \\
   && \delta\lambda_h = -\frac{1}{2v_h^3}\left(v \frac{d^2V^{\rm CW}}{d h^2} -\frac{dV^{\rm CW}}{d h} \right)\Big|_{h=v_h,s=0}.
\end{eqnarray}
For notational simplicity, we use the shorthand $V^{\rm CW}\equiv V^{\rm CW}_{1-\rm loop}(h,s)$.

\section{Loop functions and couplings}\label{app:couplings}

The loop function for χ running in the loop, corresponding to the left diagram in Fig.\,\ref{fig:one-loopDD}, is evaluated using the \texttt{Package-X} package~\cite{Patel:2015tea} and is given by
\begin{eqnarray}
&&G_\chi(m_S,m_\chi)=\frac{m_\chi \left(2m_\chi^2-m_S^2\right)\,
\mathrm{DiscB}\!\left(m_\chi^2,m_\chi,m_S\right)}
{m_S^2\left(4m_\chi^2-m_S^2\right)}
+\frac{1}{2m_\chi}\ln\!\left(\frac{m_\chi^2}{m_S^2}\right)\nonumber\\
&&+2\Bigg[
-\frac{1}{2m_\chi^2}
+\frac{\left(2m_\chi^4-4m_\chi^2m_S^2+m_S^4\right)
\mathrm{DiscB}\!\left(m_\chi^2,m_\chi,m_S\right)}
{2m_\chi^2m_S^2\left(4m_\chi^2-m_S^2\right)}\nonumber\\
&&+\frac{\left(2m_\chi^2-m_S^2\right)}
{4m_\chi^4}
\ln\!\left(\frac{m_\chi^2}{m_S^2}\right)
\Bigg]
m_\chi,
\end{eqnarray}
with 
\begin{eqnarray}
    \mathrm{DiscB}\!\left(m_\chi^2,m_\chi,m_S\right) = \frac{\sqrt{m_S^2\left(m_S^2-4m_\chi^2\right)}}{m_\chi^2}
\ln\!\left(
\frac{m_S^2+\sqrt{m_S^2\left(m_S^2-4m_\chi^2\right)}}
{2m_\chi m_S}
\right).
\end{eqnarray}
Using the \texttt{Package-X} package, the loop function when $N$ is in the loop (right diagram of Fig.\,\ref{fig:one-loopDD}) can be written as
\begin{eqnarray}
&&G_N\left(m_S,m_\chi,m_R\right)=-\frac{m_R\left(m_\chi^2+m_R^2-m_S^2\right)\,\mathrm{DiscB}\!\left(m_\chi^2,m_R,m_S\right)}
{\lambda\!\left(m_\chi^2,m_R^2,m_S^2\right)}
+\frac{m_R}{2m_\chi^2}\ln\!\left(\frac{m_R^2}{m_S^2}\right)\nonumber\\
&&+2\Bigg[
-\frac{1}{2m_\chi^2}
-\frac{\left(m_\chi^4+m_R^4-2m_\chi^2m_S^2-2m_R^2m_S^2+m_S^4\right)
\,\mathrm{DiscB}\!\left(m_\chi^2,m_R,m_S\right)}
{2m_\chi^2\,\lambda\!\left(m_\chi^2,m_R^2,m_S^2\right)}\nonumber\\
&&+\frac{\left(m_\chi^2+m_R^2-m_S^2\right)}
{4m_\chi^4}
\ln\!\left(\frac{m_R^2}{m_S^2}\right)
\Bigg]m_\chi,
\end{eqnarray}
with 
\begin{equation}
\mathrm{DiscB}\!\left(m_\chi^2,m_R,m_S\right) =\frac{\sqrt{\lambda\!\left(m_\chi^2,m_R^2,m_S^2\right)}}{m_\chi^2}
\ln\!\left(
\frac{-m_\chi^2+m_R^2+m_S^2+\sqrt{\lambda\!\left(m_\chi^2,m_R^2,m_S^2\right)}}
{2m_R m_S}
\right),
\end{equation}
where the \text{K\"all\'en} function
\begin{equation}
\lambda\!\left(m_\chi^2,m_R^2,m_S^2\right)
=
m_\chi^4+m_R^4+m_S^4
-2m_\chi^2m_R^2
-2m_\chi^2m_S^2
-2m_R^2m_S^2.
\end{equation}

\subsection*{Yukawa coupling factor in $\mathbb{Z}_3$ breaking era}
For the case of $y_2=y_3$, the Yukawa couplings of the CP-even scalar with $P_i$
\begin{eqnarray}
&&\bar{P}_1 P_1 s \longrightarrow -i\, (U^T Y U )_{11},\quad \bar{P}_2 P_2 s \longrightarrow -i\, (U^T Y U )_{22},\nonumber \\
&& \bar{\chi_2} \chi_2 s \longrightarrow i\,\frac{y_1}{\sqrt{2}},\quad \qquad \bar{P}_1 P_2 s \longrightarrow - 2i\, (U^T Y U )_{12},
\end{eqnarray}
with the Yukawa and the rotational matrix corresponding to the CP-even scalars: 
\begin{equation} Y=
\begin{pmatrix}
    \frac{y_1}{\sqrt{2}}& \frac{y_2}{2} \\[0.5mm]
    \frac{y_2}{2} &0
\end{pmatrix},\quad  U(T) = \begin{pmatrix}
    \cos\theta(T)& \sin\theta(T) \\[0.5mm]
    -\sin\theta(T) & \cos\theta(T)
\end{pmatrix}.
\end{equation}
The Yukawa couplings of the CP-odd scalar with $P_i$:
\begin{eqnarray}
&&\bar{P}_1 \chi_2 \zeta \longrightarrow  i\,\left( \sqrt{2}y_1 \cos\theta(T)+ y_2 \sin\theta(T) \right),\nonumber \\
&& \bar{P}_2 \chi_2 \zeta \longrightarrow  i\,\left( \sqrt{2}y_1 \sin\theta(T)- y_2 \cos\theta(T) \right)
\end{eqnarray}

\subsection*{Scalar vertices in $\mathbb{Z}_3$ breaking era}
\begin{eqnarray}
 &&  s\,s\,s\,s \rightarrow - 6 i\, \lambda_s,\quad \zeta\,\zeta\,\zeta\,\zeta \rightarrow - 6 i\, \lambda_s,\quad H^\dagger H\, s\,s \rightarrow -i\,\lambda_{hs},\quad H^\dagger H\,\zeta\,\zeta\rightarrow -i\,\lambda_{hs}\nonumber \\
    && s\,s \,\zeta\,\zeta \rightarrow - 2i\, \lambda_s,\quad s\,s\,s\rightarrow -2i\,(3 v_s(T)\lambda_s +  \mu_3),\quad s\, \zeta\,\zeta \rightarrow -2i\,\left(v_s(T)\lambda_s -  \mu_3\right),\nonumber\\
    && H^\dagger H\, s \rightarrow -i\, v_s(T)\lambda_{hs}.
\end{eqnarray}